\documentclass{aa}
\input {epsf}
\usepackage{graphicx}
\input psfig.sty

\begin{document}

\title{XMM observation of M~87}
\subtitle{II. Abundance structure of the interstellar and intergalactic medium}

\offprints{K. Matsushita, \\email: matusita@xray.mpe.mpg.de}
\date{Received ; accepted  }
\author{Kyoko Matsushita, Alexis Finoguenov and Hans B\"ohringer }
\authorrunning{Kyoko Matsushita et al.}
\institute{Max-Planck-Institut f\"ur Extraterrestrial Physik, D-85748 Garching, 
Germany}

\abstract{
 Based on a  detailed study of the temperature structure of 
the intracluster medium in the halo of M~87, abundance profiles 
of 7 elements, O, Mg, Si, S, Ar, Ca, and Fe are derived.
In addition, abundance ratios are derived from the ratios of line
 strengths, whose temperature dependences are small
 within the temperature range of the ICM of M~87.
The abundances of Si, S, Ar, Ca and Fe show  strong decreasing gradients outside
 2$'$ and become nearly constant  within the radius at $\sim1.5$ solar.
The Fe/Si ratio is determined to be 0.9 solar with no radial gradient.
In contrast, the O abundance is less than a half of the Si abundance at the
 center and has a  flatter gradient.
The Mg abundance is $\sim$1 solar within 2$'$, which is close to stellar
  abundance within the same radius.
The O/Si/Fe pattern of M~87 is located at the simple extension  of that of Galactic stars.
The observed Mg/O ratio is about 1.25 solar, which is also the same
ratio as for Galactic stars. 
The O/Si/Fe ratio indicates that the SN Ia contribution to Si and Fe becomes important
 towards the center and  SN Ia products have
 similar abundances of Si and Fe at least around M~87, which may reflect
 dimmer SN Ia observed in old stellar systems.
The S abundance is similar to the Si abundance at the center, but has a
steeper gradient.
This result suggests that the S/Si ratio of SN II products is much
smaller than the solar ratio.
}
\maketitle

\section{Introduction}

The intracluster medium (ICM) contains a large amount of metals, which are
mainly synthesized in early-type galaxies (e.g. Arnaud et al. 1992; Renzini et
 al. 1993).
Thus,  abundances of the metals are  tracers of chemical evolution of
galaxies and clusters of galaxies.

Based on the  Si/Fe ratio observed with ASCA,  a discussion on  contributions from SN Ia
and SN II to the metals has commenced.
In a previous  nucleosynthesis model of SN Ia, the Fe abundance is much larger
than the Si abundance in the ejecta of SN Ia (Nomoto et al. 1984).
Observations of metal poor Galactic stars indicate that average
products of SN II have a factor of 2--3 larger abundance of
$\alpha$-elements than Fe (e.g. Edvardsson et al. 1993;  Nissen et
al. 1994; Thielemann et al. 1996),
 although this ratio may depend on the initial mass function (IMF) of stars.
From elemental abundance ratios in the ICM of four clusters of galaxies observed with ASCA, 
Mushotzky et al. (1996) suggested that the metals in the ICM were mainly
produced by SN II.
Since stars in early-type galaxies in galaxy clusters  are very old
(e.g. Stanford et al. 1995; Kodama et al. 1998),  this means that most of the metals in the ICM
are produced through  star formations at high $z$.
Fukazawa et al. (1998) systematically studied 40 nearby clusters and
 found that the Si/Fe ratio is lower among 
the low-temperature clusters, which indicates that SNe Ia products are also 
important among these clusters.
From the observed radial dependence of the abundances, Finoguenov et al. 
(2000) found that the SNe II ejecta have been widely distributed in the ICM.

In the ICM, 
around a cD galaxy, the contribution of metals from the galaxy becomes important.
Fukazawa et al. (2000) found a central increment of Fe and Si abundances
around cD galaxies, which is  due to SNe Ia from the cD galaxies.
The  supply of metals by stellar mass loss must be also considered (e.g. Matsushita et al. 2000)

In addition to the Si and Fe abundances,  
the XMM-Newton observatory enables us to
obtain $\alpha$-element abundances such as for O and Mg, which are not
synthesized by SN Ia.
B\"ohringer et al. (2001) and Finoguenov et al. (2002) analyzed annular spectra
of M~87 observed by  XMM-Newton, and found a
flatter abundance gradient of O  compared to  steep gradients
of   Si, S, Ar,  Ca, and Fe.
The stronger abundance increase of Fe compared to that of O 
indicates an enhanced SN Ia contribution in the central regions (in disagreement with
our findings, Gastaldello and Molendi (2002) claim a similar gradient for
O and Fe, which would not support this conclusion).
The  observed similar abundance gradients of Fe and Si and the large Si/O ratio at
the center imply a significant contribution of Si by SN Ia that is a
larger Si/Fe abundance ratio by SN Ia than obtained by the classical model of
Nomoto et al. (1984).
The larger Si/Fe ratio and the implication that the Si/Fe ratio supplied
by  SN Ia may change with radius in the M~87 halo may indicate a
diversity of SN Ia explosions also reflected in a diversity of the light
curves as discussed by Finoguenov et al. (2002).
A similar abundance pattern of O, Si and Fe is observed around center
of A 496 (Tamura et al. 2001).

A prerequisite of the abundance determination is a precise knowledge of
the temperature structure of the ICM. 
Based on the XMM-Newton observation of M~87, Matsushita et al. (2002; hereafter
Paper I) found that  the intracluster medium  has a single phase structure 
locally, except for the regions associated with  radio jets and lobes
(B\"ohringer et al. 1995; Belsole et al. 2001), where
       there is an additional $\sim$ 1 keV temperature component.
The signature of gas cooling below 0.8 keV to zero temperature is not observed
as expected for a cooling flow (e.g Fabian et al. 1984).
The fact that the thermal structure of the intracluster medium is fairly
simple and the plasma is almost locally isothermal facilitates the
abundance determination enormously and helps to make the spectral
modeling almost unique.

In this paper, based on the detailed study of the temperature structure,
abundances of various elements are discussed. 
In section 2, we summarize the observation and data preperation.
Sections 3 and 4 deal with the MEKAL (Mewe et al. 1995, 1996; Kaastra
1992; Liedahl et al. 1994) and the APEC model (Smith et al. 2001) fits to the
deprojected spectra, and in section 5, 
abundance ratios are determined directly from the line ratios,
considering their temperature dependences.
In section 6, we evaluate an effect of  resonant line scattering.
Section 7 gives a discussion of the obtained results, and Section 8
summarizes the paper.
 We adopt for the solar abundances the values given by Feldman
(1992), where the solar Ar and Fe abundances relative to H are
4.47$\times10^{-6}$ and 3.24$\times10^{-5}$  in number, respectively.
These values are different 
from the ``photospheric'' values of Anders \& Grevesse (1989), where the
solar  Ar and Fe abundances are 3.63$\times10^{-6}$ and 4.68$\times10^{-5}$, respectively.
The solar abundances of the other elements are consistent with each other.
Unless otherwise specified,  we use 90\% confidence error regions.

\section{Observation}

M~87 was observed with  XMM-Newton on June 19th, 2000.
The effective exposures of the EPN and the EMOS are 30ks and 40ks, respectively.
The details of the analysis of  background subtraction, vignetting
correction and deprojection technique are described in Paper I.
When accumulating spectra, 
we used a spatial filter, excluding those regions where the brightness is larger by 15 \% than
the  azimuthally  averaged value in order to excise the soft emission
around the radio structures (Figure 1 and 6 in Paper I), although it is
not fully excluded within $2'$,
due to its complicated structure and the limited spatial resolution of the XMM telescope.

For the EPN spectra, 
we employ the response matrix corresponding to the average distance from
the readout for each accumulated region, epn\_fs20\_sY$i$\_thin.rmf from March 2001.
Here, $i$ reflects the distance from the readout-node, which affects the energy resolution.
For the EMOS data,  we used m1\_thin1v9q19t5r5\_all\_15.rsp from June 2001.
The spectral analysis uses the XSPEC\_v11.1 package.
We fitted  the EPN and EMOS data in the spectral range of 0.5 to 10 keV.
For outside 8$'$, the energy range between 7.5 and 8.5 keV of the EPN spectra is ignored,
because of strong emission lines induced by particle events  (Freyberg et al. 2002).

\section{Spectral fit with the MEKAL model}

We have fitted the deprojected spectra with a single
temperature MEKAL model with photoelectric absorption, in the same way
as in Paper I.
For the outermost region, we fitted the projected (annular) spectrum.
For the spectra within 1$'$,  a power-law component is added for the
central active galactic nucleus.
We fixed  index of the power-law component to the best-fit value obtained 
from the spectrum within 0.12$'$ (Paper I) and normalized it using the
point spread function of the XMM telescope.
Abundances of C and N are fixed to be 1 solar and those of other
elements are determined separately.
Within 2$'$, we also fitted the  spectra with a two temperature MEKAL
model,  where the abundances of each element of the two 
components are assumed to have the same values,
since within this radius, there remains a small amount of the cooler component with a nearly constant
temperature of 1 keV associated with the radio lobes (Belsole et al. 2001;
Paper I).
As discussed in Paper I, the two temperature model should reflect the
actual temperature structure of the ICM, although  within 0.5$'$, the XMM
spatial resolution is not enough to resolve temperature components of
the complicated structure of the ICM.

\begin{figure}[]
\resizebox{\hsize}{!}{\includegraphics{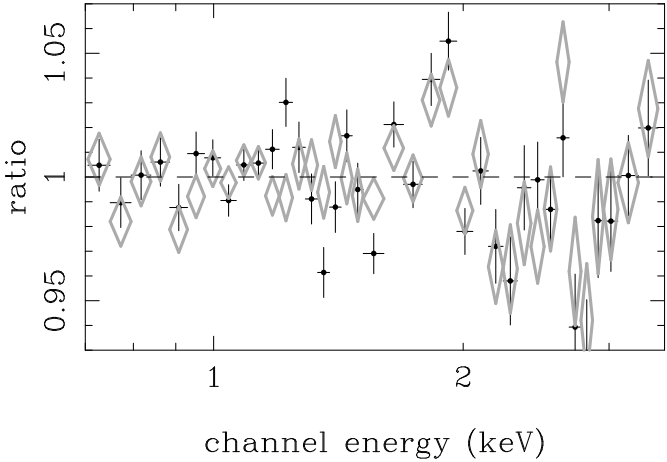}}
\caption{
 The data-to-model ratios of the deprojected spectra of the EMOS  within
 $R$=2-8$'$  fitted with the single temperature MEKAL model
 (black crosses) and  version 1.1 of the APEC model (gray diamonds).
}
\label{mos_apec}
\end{figure}

\begin{table*}
\caption{Results of spectrum fitting of the deprojected spectra with the
 single temperature and the two temperature MEKAL model. 
For the outermost region, projected spectra are used.  The
 regions with the enhanced X-ray emission associated with the radio
 structures are excluded throughout this paper.}
  \begin{tabular}[t]{rlllllllllr}
 $R$       &  kT1  & kT2 &$N_{\rm{H}}$ & O & Si & S & Ar & Ca & Fe & $\chi^2/\mu^a$\\
 (arcmin) & (keV)  & (keV) &\tiny{($10^{20}\rm{cm^{-2}}$)}
& (solar) & (solar) & (solar) &(solar)&
  (solar)& (solar) &\\
\hline
\multicolumn{11}{c}{The EMOS spectra fitted with a 1T  MEKAL model}\\
\hline
 0.00- 0.35 &1.13&  &4.1& 0.02& 0.37& 0.37& 1.01& 1.25& 0.20& 435/177 \\
 0.35- 1.00 &1.47&  &4.3& 0.00& 0.83& 0.83& 0.69& 1.84& 0.79& 563/177 \\
 1.00- 2.00 &1.56$^{+0.03}_{-0.03}$&  &3.2$^{+1.1}_{-1.1}$& 0.42$^{+0.14}_{-0.14}$& 1.29$^{+0.15}_{-0.14
}$& 1.09$^{+0.17}_{-0.15}$& 0.96$^{+0.37}_{-0.36}$& 2.66$^{+0.88}_{-0.85}$& 1.05$^{+0.08}_{-0.07}$& 177/175 \\
 2.00- 4.00 &2.05$^{+0.04}_{-0.04}$& &1.3$^{+0.6}_{-0.6}$& 0.44$^{+0.09}_{-0.09}$& 1.24$^{+0.10}_{-0.10
}$& 1.12$^{+0.11}_{-0.10}$& 0.71$^{+0.20}_{-0.20}$& 1.33$^{+0.39}_{-0.39}$& 1.07$^{+0.06}_{-0.06}$& 172/175 \\
 4.00- 8.00 &2.30$^{+0.04}_{-0.04}$& &1.4$^{+0.4}_{-0.4}$& 0.45$^{+0.07}_{-0.07}$& 0.92$^{+0.07}_{-0.07
}$& 0.71$^{+0.08}_{-0.07}$& 0.42$^{+0.14}_{-0.16}$& 1.01$^{+0.28}_{-0.27}$& 0.79$^{+0.04}_{-0.04}$& 199/175 \\
 8.00-11.30 &2.51$^{+0.08}_{-0.08}$& &1.9$^{+0.7}_{-0.7}$& 0.37$^{+0.12}_{-0.11}$& 0.64$^{+0.11}_{-0.10
}$& 0.35$^{+0.12}_{-0.11}$& 0.37$^{+0.23}_{-0.24}$& 0.72$^{+0.42}_{-0.43}$& 0.62$^{+0.06}_{-0.06}$& 243/175 \\
11.30-13.50 &2.72$^{+0.06}_{-0.05}$& &0.6$^{+0.4}_{-0.4}$& 0.32$^{+0.07}_{-0.07}$& 0.61$^{+0.07}_{-0.06
}$& 0.20$^{+0.07}_{-0.07}$& 0.14$^{+0.15}_{-0.14}$&
   0.49$^{+0.24}_{-0.27}$& 0.51$^{+0.03}_{-0.03}$& 275/175 \\
\hline
\multicolumn{11}{c}{The EMOS spectra fitted with a 2T  MEKAL model}\\
\hline
 0.00- 0.35 &1.56$^{+0.19}_{-0.16}$& 0.75$^{+0.02}_{-0.02}$ &2.3$^{+1.5}_{-1.5}$& 0.24$^{+0.30}_{-0.23}$& 1.04$^{+0.62}_{-0.3
8}$& 1.07$^{+0.63}_{-0.43}$& 0.62$^{+1.07}_{-0.62}$& 2.03$^{+2.46}_{-2.03}$& 0.85$^{+0.39}_{-0.24}$& 138/176 \\
 0.35- 1.00 &1.73$^{+0.04}_{-0.04}$& 0.91$^{+0.03}_{-0.03}$ &2.1$^{+0.8}_{-0.4}$& 0.65$^{+0.13}_{-0.12}$& 1.70$^{+0.17}_{-0.1
6}$& 1.53$^{+0.17}_{-0.16}$& 0.99$^{+0.28}_{-0.28}$& 1.38$^{+0.58}_{-0.57}$& 1.54$^{+0.12}_{-0.11}$& 227/176 \\
 1.00- 2.00 &1.68$^{+0.06}_{-0.06}$& 0.97$^{+0.15}_{-0.15}$ &2.6$^{+1.1}_{-1.2}$& 0.55$^{+0.19}_{-0.16}$& 1.56$^{+0.26}_{-0.1
9}$& 1.25$^{+0.23}_{-0.18}$& 0.96$^{+0.41}_{-0.39}$& 2.38$^{+0.92}_{-0.87}$& 1.28$^{+0.09}_{-0.12}$& 162/176 \\
\hline
\hline
\multicolumn{11}{c}{The EPN spectra fitted with a  1T
  MEKAL model}\\
\hline
 0.00- 0.35 &0.74&  &9.0& 0.19& 0.32& 0.71& 3.15& 0.40& 0.08& 255/132 \\
 0.35- 1.00 &1.44&  &4.6& 0.28& 0.86& 0.86& 0.79& 2.06& 0.81& 296/132 \\
 1.00- 2.00 &1.56$^{+0.05}_{-0.04}$&  &3.5$^{+1.1}_{-0.5}$& 0.45$^{+0.15}_{-0.14}$& 1.09$^{+0.17}_{-0.16}$& 0.99$^{+0.22}_{-0.20}$& 1.30$^{+0.56}_{-0.54}$& 1.70$^{+1.28}_{-1.25}$& 0.98$^{+0.09}_{-0.08}$& 112/125 \\
 2.00- 4.00 &1.98$^{+0.06}_{-0.06}$&  &1.5$^{+0.7}_{-0.7}$& 0.40$^{+0.11}_{-0.10}$& 1.05$^{+0.14}_{-0.13}$& 0.95$^{+0.16}_{-0.15}$& 0.81$^{+0.35}_{-0.35}$& 1.63$^{+0.69}_{-0.68}$& 0.96$^{+0.08}_{-0.07}$& 125/125 \\
 4.00- 8.00 &2.28$^{+0.06}_{-0.06}$&  &1.4$^{+0.5}_{-0.6}$& 0.28$^{+0.09}_{-0.08}$& 0.71$^{+0.10}_{-0.10}$& 0.53$^{+0.13}_{-0.12}$& 0.21$^{+0.27}_{-0.21}$& 1.01$^{+0.50}_{-0.50}$& 0.69$^{+0.05}_{-0.05}$&  93/125 \\
 8.00-11.30 &2.48$^{+0.10}_{-0.11}$& &1.7$^{+0.8}_{-0.9}$& 0.29$^{+0.16}_{-0.14}$& 0.58$^{+0.19}_{-0.17}$& 0.28$^{+0.21}_{-0.23}$& 0.41$^{+0.46}_{-0.41}$& 0.95$^{+0.81}_{-0.84}$& 0.53$^{+0.09}_{-0.07}$&  85/125 \\
11.30-16.0 &2.54$^{+0.07}_{-0.07}$&  &0.5$^{+0.5}_{-0.4}$& 0.26$^{+0.07}_{-0.07}$& 0.37$^{+0.09}_{-0.08
}$& 0.33$^{+0.12}_{-0.11}$& 0.20$^{+0.24}_{-0.20}$& 0.38$^{+0.39}_{-0.38}$& 0.41$^{+0.04}_{-0.04}$& 190/131 \\
\hline
\multicolumn{11}{c}{The EPN spectra fitted with a  2T  MEKAL model}\\
\hline
 0.00- 0.35 &1.73$^{+0.16}_{-0.18}$& 0.73$^{+0.02}_{-0.02}$ &0.0$^{+1.4}_{0.0}$& 0.48$^{+0.39}_{-0.27}$& 1.23$^{+0.57}_{-0.49}$& 1.02$^{+0.74}_{-0.61}$& 1.27$^{+1.77}_{-1.27}$& 1.26$^{+3.31}_{-1.26}$& 1.08$^{+0.32}_{-0.32}$& 133/123 \\
 0.35- 1.00 &1.73$^{+0.05}_{-0.05}$& 0.87$^{+0.03}_{-0.03}$  &0.9$^{+1.0}_{-0.9}$& 0.58$^{+0.14}_{-0.13}$& 1.47$^{+0.26}_{-0.23}$& 1.56$^{+0.29}_{-0.25}$& 0.96$^{+0.48}_{-0.47}$& 1.65$^{+0.96}_{-0.93}$& 1.55$^{+0.18}_{-0.17}$& 152/123 \\
 1.00- 2.00 &1.77$^{+0.09}_{-0.08}$& 1.00 (fix) &0.8$^{+1.4}_{-0.6}$& 0.69$^{+0.23}_{-0.20}$& 1.73$^{+0.41}_{-0.35}$& 1.47$^{+0.42}_{-0.36}$& 1.60$^{+0.78}_{-0.71}$& 1.38$^{+1.52}_{-1.38}$& 1.49$^{+0.26}_{-0.23}$&  90/123 \\
\hline
 \end{tabular}
 \\ $^a$ Degrees of freedom
\end{table*}

\begin{figure*}[]
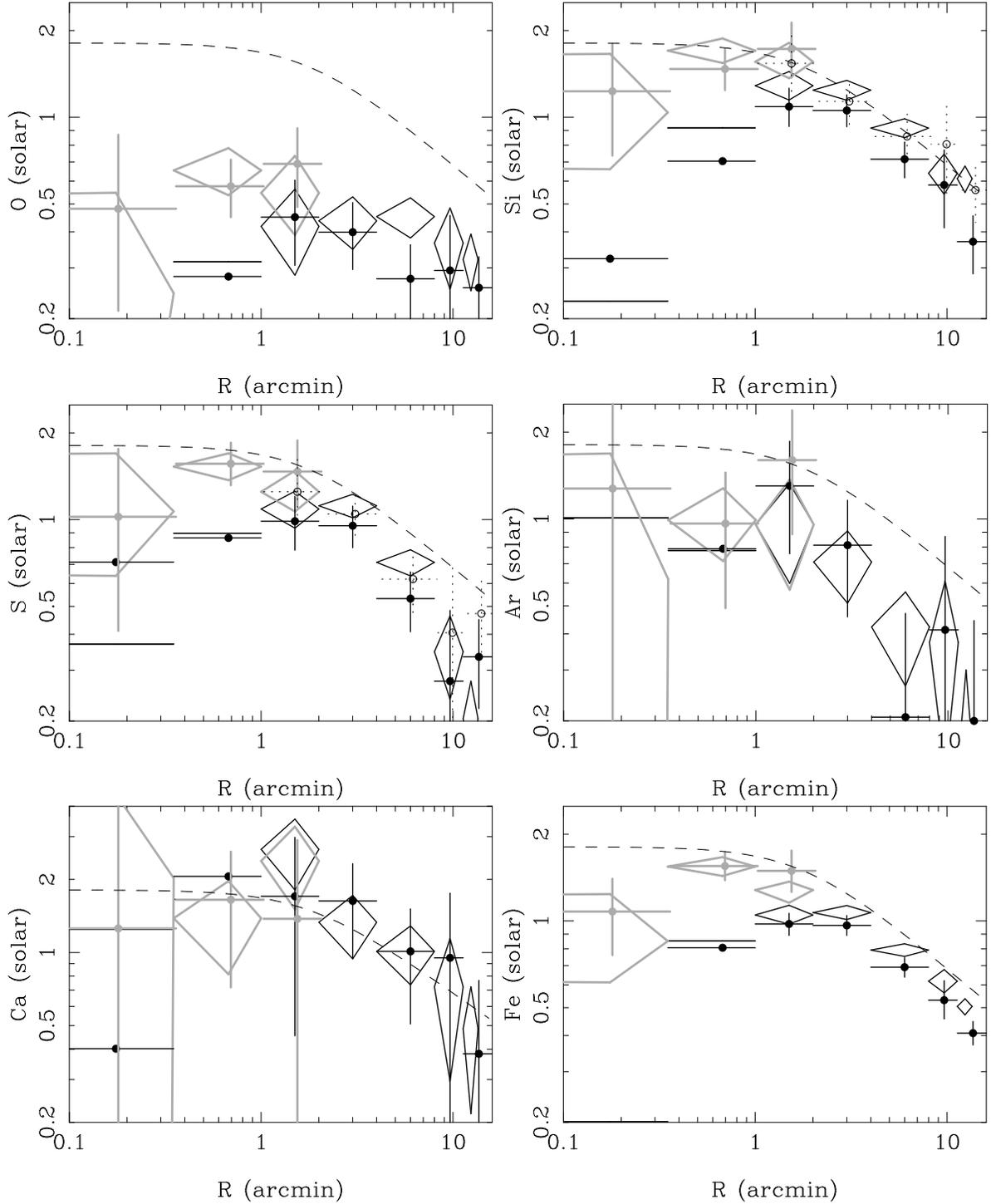

\centerline{
\resizebox{7.8cm}{!}{\includegraphics{matsushita_fig2a.ps}}
\resizebox{7.8cm}{!}{\includegraphics{matsushita_fig2b.ps}}
}
\centerline{
\resizebox{7.8cm}{!}{\includegraphics{matsushita_fig2c.ps}}
\resizebox{7.8cm}{!}{\includegraphics{matsushita_fig2d.ps}}
}
\centerline{
\resizebox{7.8cm}{!}{\includegraphics{matsushita_fig2e.ps}}
\resizebox{7.8cm}{!}{\includegraphics{matsushita_fig2f.ps}}
}
\caption{
Abundance profiles of O, Si, S, Ar, Ca and Fe using the single
 temperature (black) and the two temperature (gray) MEKAL model
 from the deprojected data of the EMOS (diamonds) and the EPN (closed circles).
The open circles  are the Si and S abundances derived from  the EPN spectra above 1.7
 keV, when O, Ne, Mg, Fe and Ni abundances are fixed to be the best fit values
 derived from the EMOS data of the whole energy band.
The dashed lines correspond to the best fit  relation of
the  deprojected Si profile of the  EMOS.
}
\label{depro_abund}
\end{figure*}

The results are summarized in Table 1 and Figures 1--4.
Representative  spectra fitted with the MEKAL model are shown in 
  Figure 2 and 7 of Paper I and Figure \ref{mos_apec}. 
There remain small discrepancies between the data and the best fit
model.
A residual structure exists at 0.8--1 keV of the
deprojected  spectra of the EPN of  $R>$2$'$, when fitted with the
single temperature MEKAL model, while the EMOS data have no such
residual (Figure 2 of Paper I).
Here, $R$ is the 3-dimensional radius.
This problem will be discussed in the next paragraph.
The Fe-L/Mg-K structure around 1.3 keV is not fitted well (Figure \ref{mos_apec}).
Therefore, we do not show Mg abundances here and will discuss this
in more detail below (Sec. 4 and 5.5-5.6).
We also do not show Ne and Ni abundances, because their line strengths 
are much smaller than those of the Fe-L features at the same wavelength and a small uncertainty in the
Fe-L complex gives large uncertainties in abundances of these elements (Masai 1987;
Matsushita et al. 1997; 2000).
There are also small discrepancies in the continuum level around K$\alpha$ lines
of Si,  S and Ar, which will be discussed  in Sec. 5.

The derived abundances are mostly consistent between the EPN and the
EMOS, although the EPN abundances of O, Si, S and Fe are systematically smaller than the EMOS
results by 10--30\%, when fitted with a single temperature MEKAL model.
As will be shown  in Sec. 5, the strengths of K$\alpha$
lines of H-like  O, Si, S and He-like (+ Li like) Fe observed by the
EPN and EMOS  agree within several percent when considering the difference in
the normalization between the two  detectors.
Therefore, these abundance discrepancies  should be caused by the
discrepancy at 0.8--1 keV, which may reflect uncertainties in 
an instrumental low energy tail of the strong Fe-L lines observed in EPN
 spectra, as discussed in Paper I.
In addition, the EPN is characterized by a dependence of the energy
resolution on the position on the detector.
As a result,  the observed Fe abundance of the EPN may be artificially 
decreased by these problems and
the change of the continuum level also affects the O abundance.
Those of Si and S also couple with abundances of other elements such as
O and Fe due to the effect of free-bound emission and
the increment of the strength of the Fe-L emission around the K$\alpha$
line of He-like Si.
When we fix the O, Ne, Mg, Fe and Ni abundances to the best fit values derived from the EMOS spectra of the
whole energy band and fitted the EPN spectra above 1.7 keV, we recover  consistent Si and S
abundances between the two detectors
(Figure \ref{depro_abund}).

The derived abundances of Si, S, Ar, Ca, and Fe from deprojected spectra using the
single temperature MEKAL model fit show strong negative  gradients
at  $R>1'$, and drop sharply within this radius (Table 1, Figure \ref{depro_abund}) as
already seen in the projected spectra 
(B\"ohringer et al. 2001; Finoguenov et al. 2001; Gastaldello \& Molendi
2002).
But now the absolute
values are slightly larger than those from projected spectra.
In contrast, the O abundance is  nearly constant at $R>1'$.

\begin{figure}[]
\resizebox{9cm}{!}{\includegraphics{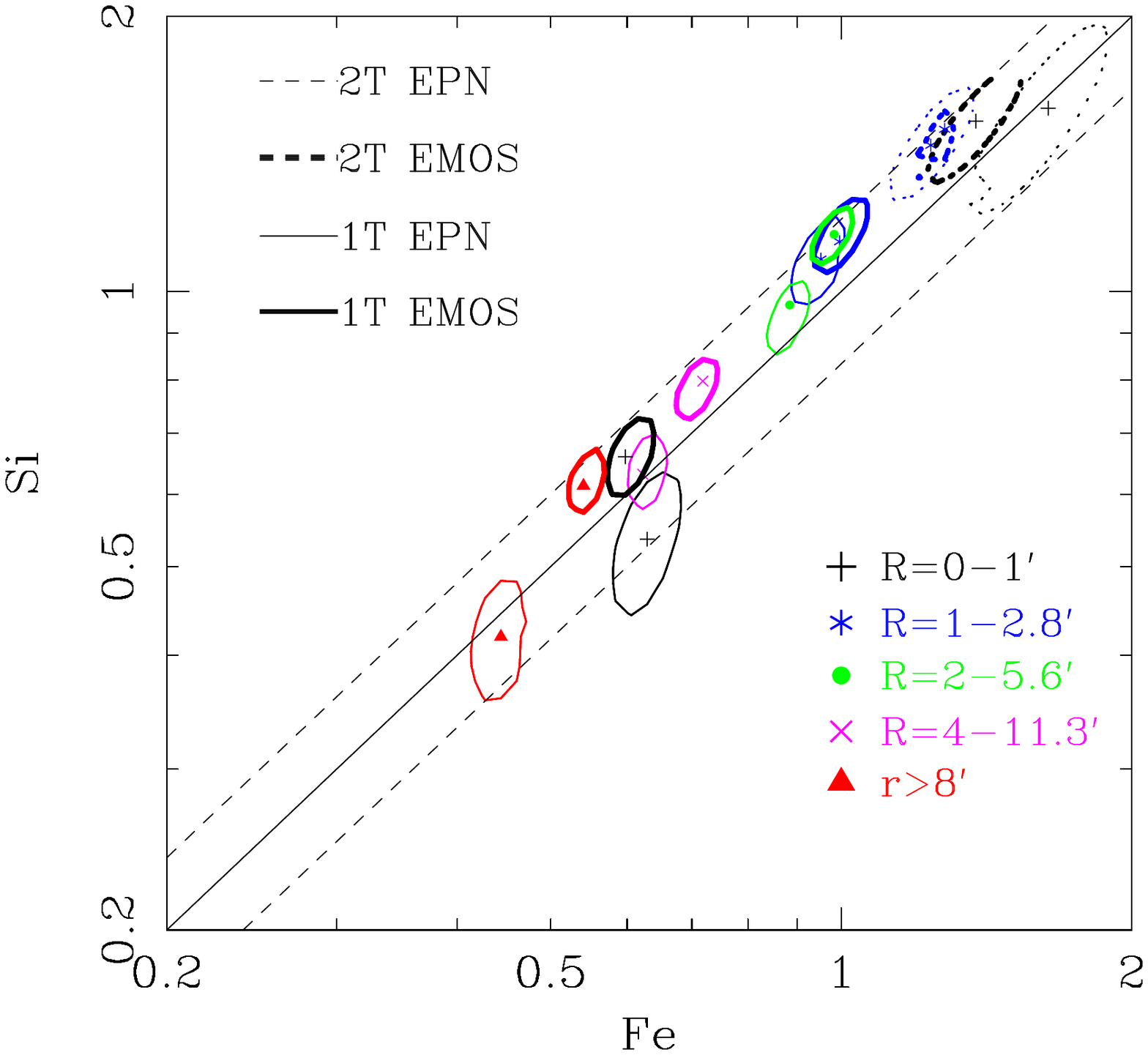}}
\resizebox{9cm}{!}{\includegraphics{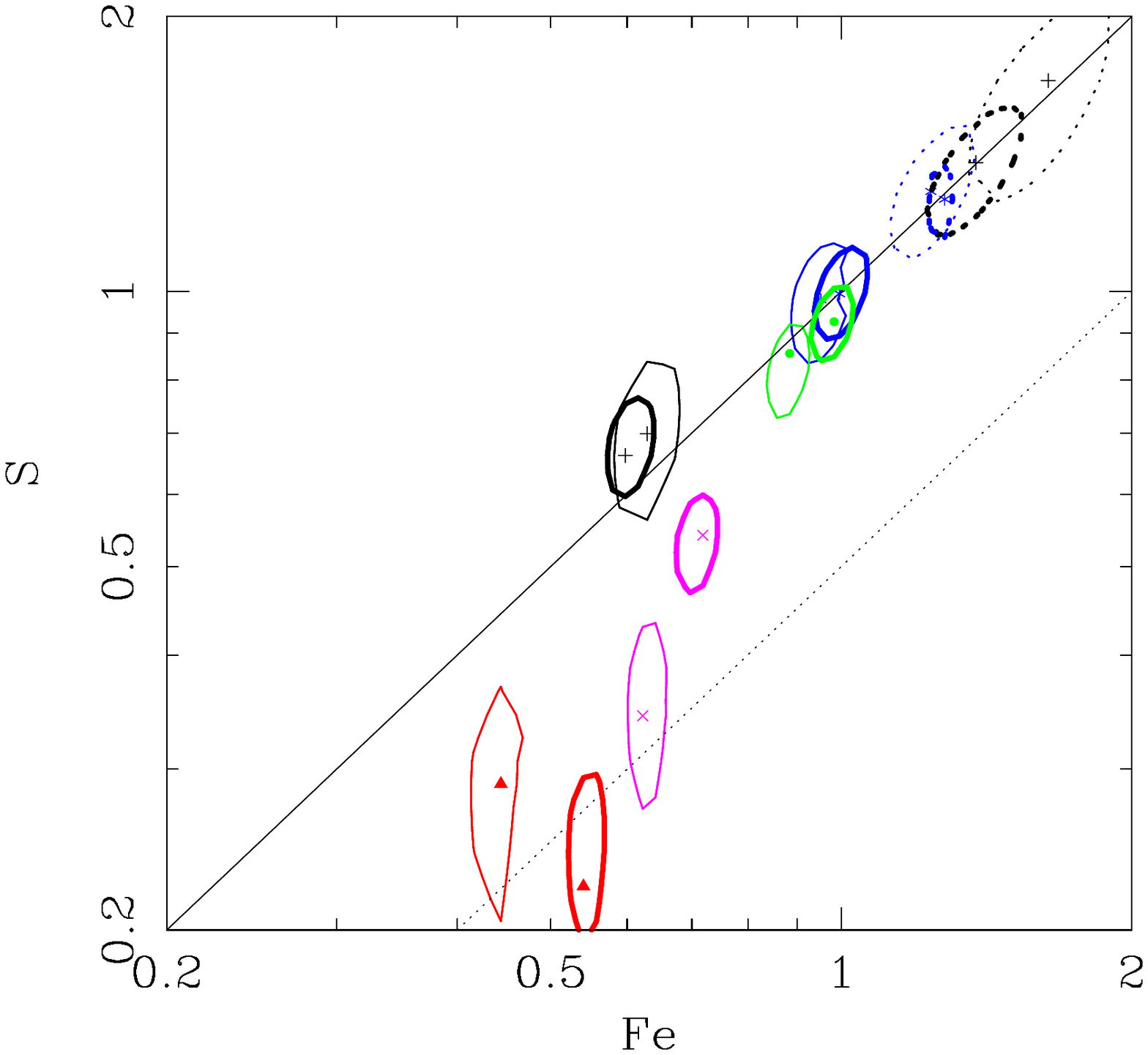}}
\caption{
Confidence contours (90\%) for the Si,  S vs.  Fe  abundance of the  deprojected data of
 the EMOS (thick lines) and the EPN (thin lines) fitted with the
 single temperature MEKAL model (solid lines) and the two temperature MEKAL
 model (dotted lines).  For the outermost region, the projected spectra are
 used. 
The solid lines indicate the condition of the solar-abundance
 ratio. The dashed lines represent the Si/Fe ratio to be  0.8 and 1.2 solar and 
the dotted line represents the S/Fe ratio to be 0.5 solar.
}
\label{abund_cont_fesi}
\end{figure}

\begin{figure}
\resizebox{9cm}{!}{\includegraphics{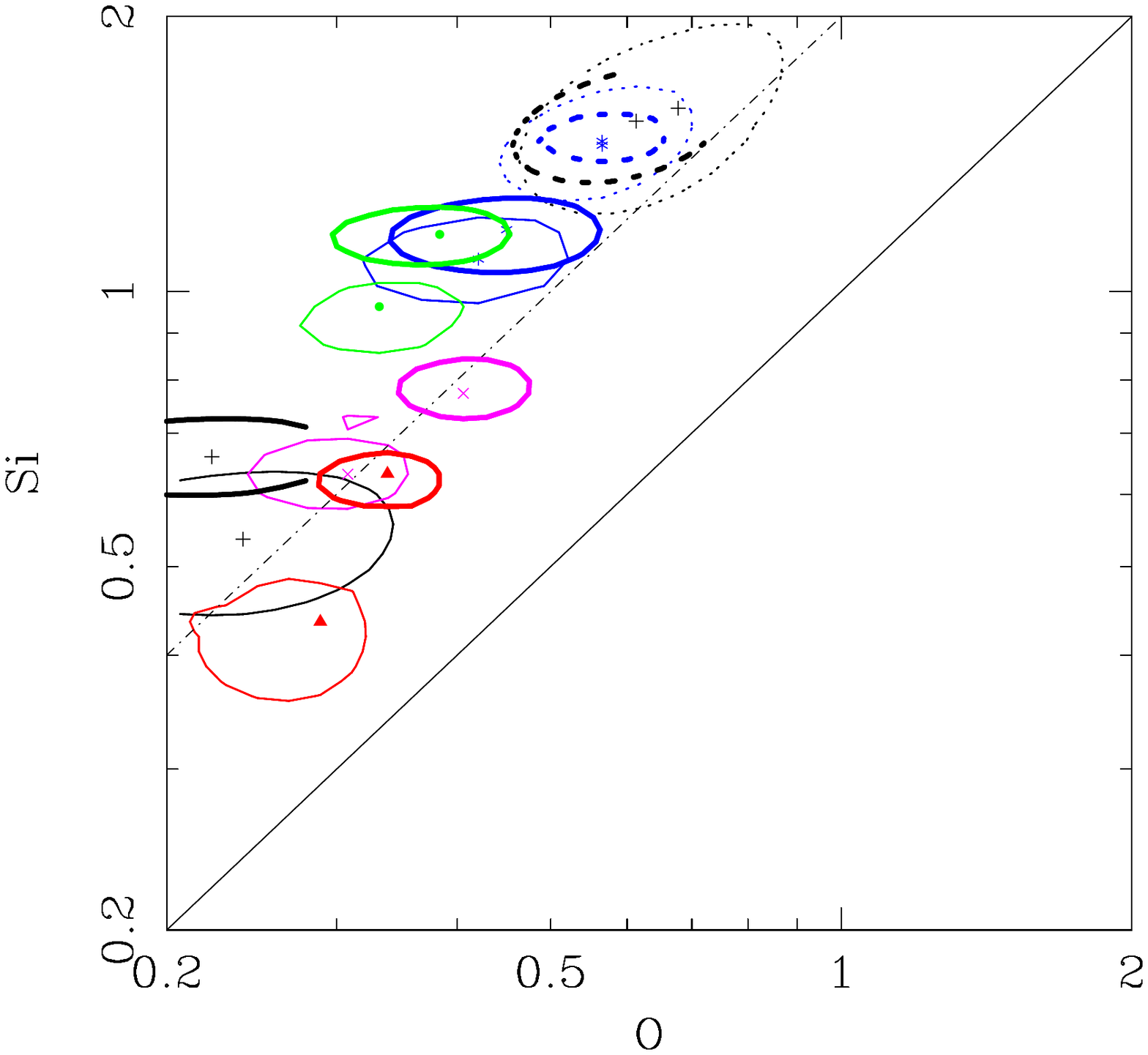}}
\resizebox{9cm}{!}{\includegraphics{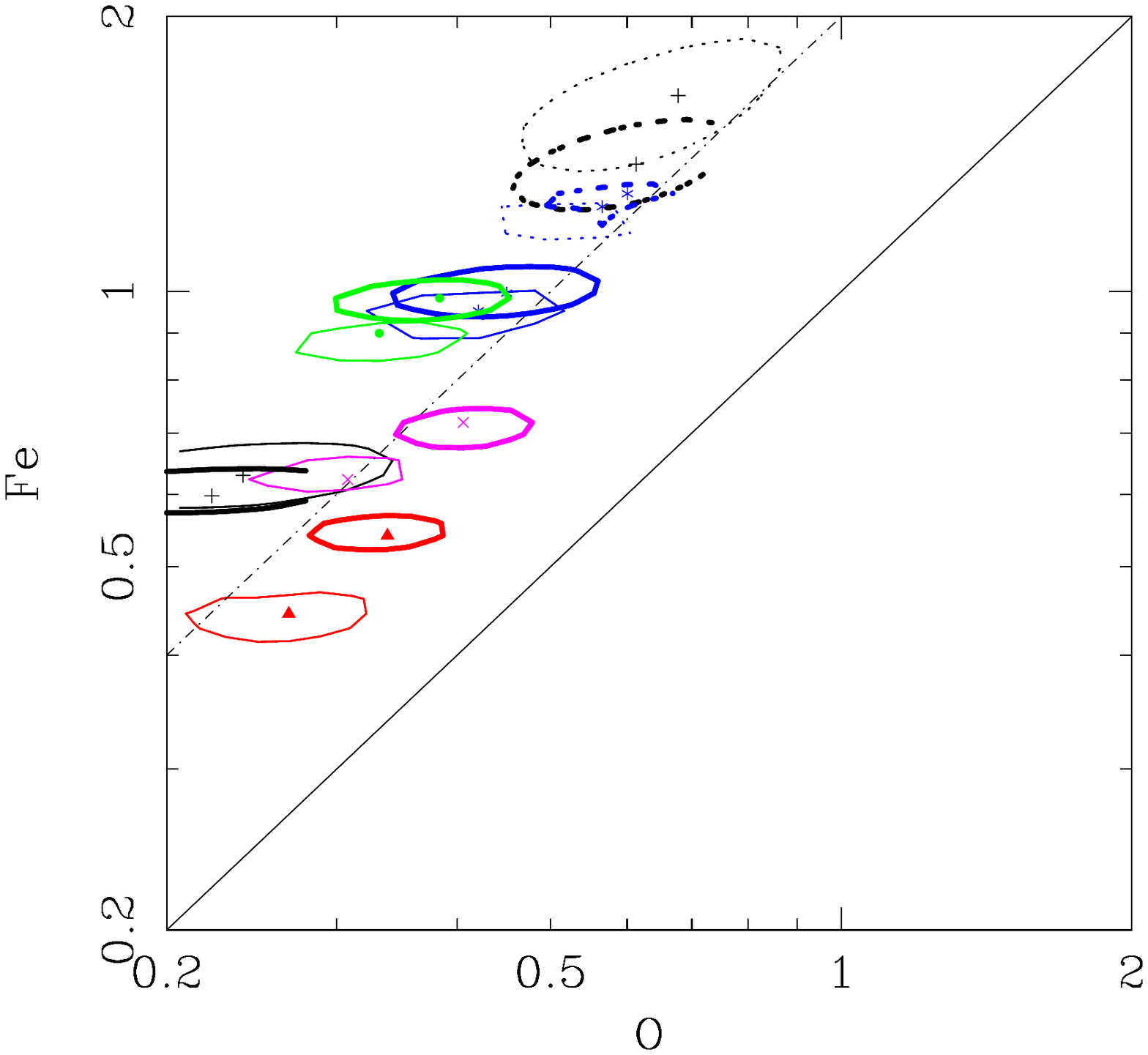}}
\caption{
90 \% confidence contours for the Si, Fe vs. O.
The meanings of the symbols are the same as in Fig. 3.
The solid lines and dot-dashed lines represent the Si or Fe to O ratio
 to be 1  and 2 solar, respectively.
}
\label{abund_cont_o}
\end{figure}

The two temperature MEKAL fit on the deprojected spectra
gives significantly larger abundances than the single temperature MEKAL
fit (Figure \ref{depro_abund}-\ref{abund_cont_o}, Table 1) 
and the central abundance drops seen in the single MEKAL fit become
very week in the results of the two temperature MEKAL fit,
although contributions of the cooler component are  small as shown in Paper I.
For example,  at $R$=0.5-1',  10 percent of the cooler component in units of emission measure
changes the Fe abundances by a factor of 2.
The reason for this is the change of the temperature of the hot
component, when the cold component is added.
 For a fixed Fe line feature this leads to a higher abundance required in the fit.
In addition, abundances of O, Si and S also increase by  a factor of 2.
The increment of the O abundance is due to the change of the temperature
structure.
 The Si and S abundances increase due to an increment of the free-bound
emission and the strength of Fe-L lines around the He-like Si line.
Therefore, considering the remaining  1 keV temperature component associated with the
radio structures within 2$'$,   abundances become nearly constant within 2$'$.

We have also tried a three temperature MEKAL model for spectra within 2$'$.
Although the derived $\chi^2$ values have slightly improved, the derived abundances
and their errors have nearly the 
same values as those  from the two temperature MEKAL model fit.


Figure \ref{abund_cont_fesi} shows the confidence contours for the Si, S and Fe abundances, derived from
the deprojected spectra.
For the outermost region, the results from the projected spectra are plotted.
The elliptical shape of the confidence contours indicates that abundance
ratios are better determined than the abundances themselves.
 The differences in the Si/Fe and S/Fe ratios between the EMOS and
EPN are considerably smaller than the abundance differences.
Furthermore, the two temperature MEKAL model fit gives  almost the same
values of the Si/Fe and S/Fe ratio with the single temperature MEKAL model fit, although absolute values
of Si, S, and Fe are largely changed.
This means that 
 the abundance ratios do not strongly depend on the temperature structure.
As a result, the Si/Fe ratio is determined to be nearly constant at 1.1$\pm$ 0.1 solar,
with no radial gradient.
In contrast, we find that the S  abundance has a steeper gradient than
Fe abundance.

The abundances of Si and Fe are nearly independent to the O abundance
(Figure \ref{abund_cont_o}).
However, at the center,  the two temperature MEKAL model gives the similar
values of the O/Si and O/Fe ratio as the single temperature MEKAL model.
The O/Si and O/Fe ratios are consistent between the two detectors,
although the abundances of O, Si, and Fe between the EMOS and EPN differ
by 10--30\%.
Both the Si/O and Fe/O ratios increase by a factor of 1.5 with radius. 
This result is in disagreement with  the claim by Gastaldello \& Molendi
(2002), where they derived the constant Fe/O ratio.
The background subtraction should not be a problem when determining O,
Fe, and Si abundances,
 since even at the radius of 10$'$, the background is only a few percent
 between 0.5 to 2 keV.
In addition, the background was subtracted according to
 the detailed study by Katayama et al. (2002). 
A more severe problem should be the uncertainties in the response matrix, since the O abundance was
inconsistent between the EPN and the EMOS by a factor of 2 using an old response matrix
(B\"ohringer et al. 2001), although the discrepancy becomes smaller now
as shown in this section.
 Using an old version of the response matrix and fitting the projected spectra
in the same way in  Gastaldello \& Molendi (2002), we achieve  the same results.

In summary,  from the deprojected spectra using the MEKAL model, 
 the abundances of Si, S, Ar, Ca, and Fe have steep gradients at
$R>2'$ and become nearly constant within the radius.
The Si/Fe ratio is almost constant within the field of view, while the
S/Fe ratio has a negative, and the O/Fe ratio has a positive gradient.

\section{Spectral fits with the APEC model}

A new plasma code, APEC (Smith et al. 2001), is now available in
XSPEC\_v11.
Within the energy resolution of the CCD detectors, the dominant change
from  version 1.0 to the APEC code in the XSPEC version 11.0.1 is not
in the Fe-L lines, but in the  K$\alpha$ lines.
Using  version 1.0 of the APEC code,  strengths of the  K$\alpha$ lines of
H-like ions, which are the most fundamental lines,  
decrease by $\sim$ 50 \% at 2 keV, and their temperature dependences also change.
The strength of the 6.7 keV Fe-K line  decreases,  by a factor of 2 at 1 keV and 1.3 at 4 keV.
The changes of the K$\alpha$ lines of He-like ions of Si, S, Ar and Ca are smaller.
As a result, any single temperature APEC model cannot fit the  deprojected
spectra of M~87, especially for the ratio of the Fe-L and the Fe-K as shown
in Paper I, although a single temperature MEKAL model can fit  each
deprojected spectrum at $R>2'$.

However, the strengths of K$\alpha$ lines of  version 1.1 of the APEC code in  version 11.1 of the
XSPEC became more consistent with the MEKAL ones.
Within the temperature range of the ICM of M~87, 
the largest differences except for the Fe-L structure are 20\% changes in the
strengths of the K$\alpha$ line of H-like  O and the Fe line at 6.7 keV.

\begin{figure}[]
\resizebox{8.1cm}{!}{\includegraphics{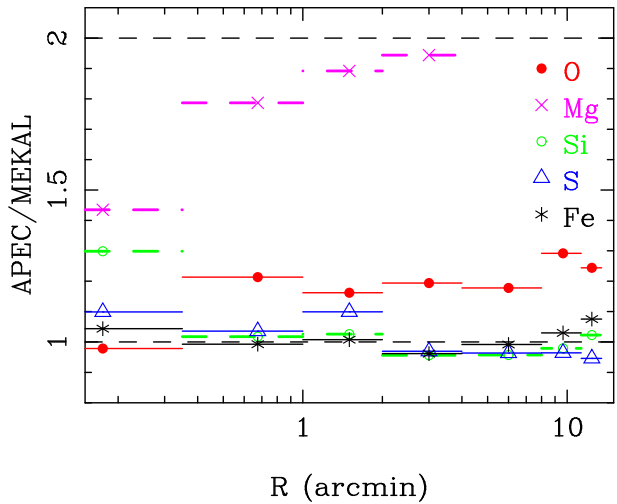}}
\caption{
 The ratios of the best fit abundances derived from the APEC model and MEKAL
 model. The two temperature model is used at $R<2'$ and the single
 temperature model is used at $R>2'$.
}
\label{ratio_apec_mekal}
\end{figure}

Using version 1.1 of the APEC code, we fitted the deprojected
spectra of the EMOS.
The results are summarized in Table 2 and Figure \ref{ratio_apec_mekal}.
In contrast to the worse $\chi^2$ values obtained from version 1.0 of the
APEC code,   version 1.1 gives  similar $\chi^2$ values compared to 
 those obtained by the MEKAL model.
For the Fe-L/Mg-K structure around $\sim 1.4$ keV, 
the APEC (v1.1) model gives a better fit  than the MEKAL model (Figure \ref{mos_apec}).
The derived temperatures and hydrogen column densities are consistent
with those from the MEKAL model.
The contribution of the lower temperature component at $R$=1--2$'$ has slightly
decreased from 4\% by the MEKAL code to 2\% by the APEC code,
although the contribution within 1$'$ is consistent with each other. 
The few percent of the soft component using the MEKAL model at $R$=1--2$'$ may
be an artifact  due to the uncertainties in the Fe-L complex.

\begin{table*}[]
\caption{The results of the spectral fits of the deprojected spectra of the EMOS using the APEC (v1.1) code }
 \begin{tabular}[t]{rcccccccccc}
 $R$       &  kT1 & kT2 & $N_{\rm{H}}$ & O  & Mg &Si & S & Fe   & $\chi^2/\mu^a$\\
 (arcmin) & (keV)  & (keV)&\tiny{($10^{20}\rm{cm^{-2}}$)}  &(solar)  &(solar)&  (solar)& (solar)& (solar) &\\
\hline 
\multicolumn{9}{c}{ a 1T  APEC model}\\\hline
0.00- 0.35 & 1.15 & & 2.5 & 0.06 & 0.0 & 0.29 & 0.41 &0.15  & 253/175\\
0.35- 1.00 & 1.48 & & 2.4 & 0.97 & 1.17 & 1.80 & 1.79 &1.37 & 583/175\\
 1.00- 2.00 &1.59$^{+0.01}_{-0.03}$& &2.8$^{+1.2}_{-1.0}$& 0.56$^{+0.17}_{-0.08}$& 0.87$^{+0.17}_{-0.17}$& 1.47$^{+0.10}_{-0.08}$& 1.28$^{+0.15}_{-0.15}$& 1.15$^{+0.05}_{-0.05}$& 126/175 \\
 2.00- 4.00 &2.01$^{+0.02}_{-0.01}$&&2.1$^{+0.3}_{-0.3}$& 0.54$^{+0.06}_{-0.09}$& 0.67$^{+0.12}_{-0.11}$& 1.19$^{+0.07}_{-0.06}$& 1.09$^{+0.08}_{-0.08}$& 1.03$^{+0.03}_{-0.02}$& 181/175 \\
 4.00- 8.00 &2.28$^{+0.02}_{-0.03}$& &2.0$^{+0.2}_{-0.4}$& 0.52$^{+0.08}
_{-0.05}$& & 0.87$^{+0.06}_{-0.05}$& 0.68$^{+0.07}_{-0.06}$& 0.78$^{+0.02
}_{-0.02}$& 210/175 \\
 8.00-11.30 &2.48$^{+0.03}_{-0.03}$&&2.2$^{+0.7}_{-0.7}$& 0.46$^{+0.1
5}_{-0.11}$& & 0.65$^{+0.08}_{-0.12}$& 0.36$^{+0.10}_{-0.14}$& 0.64$^{+0.
06}_{-0.05}$& 248/175 \\
11.30-13.50 &2.74$^{+0.05}_{-0.07}$&  &0.8$^{+0.5}_{-0.5}$& 0.38$^{+0.13}
_{-0.08}$& & 0.62$^{+0.07}_{-0.06}$& 0.19$^{+0.09}_{-0.08}$& 0.53$^{+0.03
}_{-0.01}$& 305/175 \\
\hline
\multicolumn{9}{c}{ a 2T  APEC model}\\\hline
 0.00- 0.35 &1.41$^{+0.27}_{-0.43}$& 0.78$^{+0.03}_{-0.02}$ &2.4$^{+1.1}_{-1.5}$& 0.26$^{+0.29}
_{-0.26}$& 0.75$^{+0.87}_{-0.31}$& 1.14$^{+0.96}_{-0.24}$& 1.18$^{+0.63}_{-0.42}$& 0.96$^{+0.10
}_{-0.10}$& 138/174 \\
 0.35- 1.00 &1.78$^{+0.06}_{-0.05}$& 1.05$^{+0.01}_{-0.02}$ &2.0$^{+0.8}_{-0.6}$& 0.80$^{+0.13}
_{-0.14}$& 1.10$^{+0.14}_{-0.18}$& 1.78$^{+0.09}_{-0.16}$& 1.61$^{+0.11}_{-0.16}$& 1.58$^{+0.06
}_{-0.14}$& 190/174 \\
 1.00- 2.00 &1.63$^{+0.01}_{-0.01}$& 1.00 (fix) &2.5$^{+0.6}_{-1.2}$& 0.64$^{+0.2
3}_{-0.20}$& 1.00$^{+0.32}_{-0.27}$& 1.61$^{+0.27}_{-0.23}$& 1.38$^{+0.26}_{-0.22}$& 1.26$^{+0.
17}_{-0.15}$& 125/174 \\
\hline
 \end{tabular}
 \\ $^a$ Degrees of freedom
\end{table*}

Figure \ref{ratio_apec_mekal} compares the abundances derived from  the APEC and the MEKAL model.
The large abundance changes seen in  APEC version 1.0 (Paper I)
are not derived using the new APEC code,  version 1.1. 
  For Si, S, and  Fe, we find  consistent results within several percent.
The O abundances from the APEC code are systematically larger than those from the
MEKAL code by 20\%, which is close to 
the difference of the line strength of the K$\alpha$ line of H-like O  between
the MEKAL and APEC model.

 In contrast, the derived Mg abundances differ by a factor of 2
between the two codes.
Since the APEC model can fit the Mg-K/Fe-L structure at $\sim$1.4 keV quite well, 
the APEC code may be better to derive  a Mg abundance.
There is another problem due to the  strong instrumental line  of the EMOS, which is
located at a similar energy to the Mg line.
At $r$=8--13.5$'$, the strength of the instrumental line is a factor of
10 larger than the Mg line of the ICM, and the uncertainty affects  the spectrum at
$R$=4--8$'$ through the deprojection. 
Here, $r$ is the projected radius from the center.
Therefore, the Mg abundances are shown only within 4$'$.
The derived Mg abundance is $\sim 1$ solar at $R<2'$ and shows a radial
gradient outside this radius (Table 2).

 The observed values of the Ne abundance derived from the APEC model are
also a factor of 1.3 larger than those derived from the MEKAL model.
For the Ni abundance, the APEC code gives  60\% of the values from
the MEKAL code at $R>2'$, but within $R<2'$,  both codes give 
similar values. 
In contrast to the Mg lines, the Ni-L lines and the Ne-K lines cannot be
distinguished in the spectra of M~87. 
Therefore, we do not present Ni and Ne abundances in this paper.

In summary, the new APEC code provides a better fit to the Fe-L structure,
and the derived abundances of O, Si, and Fe are mostly consistent with those obtained from the
MEKAL code.
The Mg abundances are derived to be $\sim$ 1 solar at $R<2'$, since 
the Fe-L/Mg-K structure is well fitted with the APEC code.

\section{Abundance determination using line ratios}

In this section,  we try to obtain abundance ratios of elements directly from ratios of emission lines.
When calculating the deprojected spectra,
we make an assumption that spectra beyond the field of view are the same as the outermost spectrum,
which is not valid since the temperature starts to decrease beyond the field
of view of the detector (Shibata et al. 2001).
Therefore, especially in the outer regions, there may be a bias due to this assumption,
since emission lines have a stronger dependence on temperature than  the continuum.
In addition, abundances from deprojected spectra have a larger uncertainty due to
poorer statistics than those derived from projected data.
In order to derive abundances more accurately, we need
projected spectra.

\begin{figure}[]
\resizebox{\hsize}{!}{\includegraphics{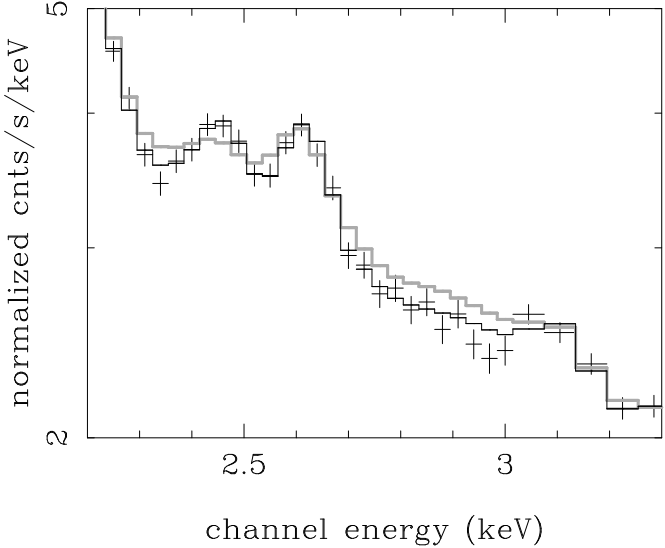}}
\caption{The projected EMOS spectrum of $r$=8.0-13.5$'$ (crosses) fitted with a single
 MEKAL model (gray) and bremstrallung and gaussians (black).
}
\label{mekalspec}
\end{figure}

However, the projected spectra consist of multi-temperature components.
As a result,  some line ratios, such as the ratio of  K$\alpha$ lines of He- and H-like S, of the
projected spectra cannot be fitted with a single temperature MEKAL model
 (Figure \ref{mekalspec}).
We must be careful in using a multi-temperature model, e.g. a two
temperature model, since emission lines and  a continuum spectrum are different
functions of temperature.
In addition,
the continuum spectra between S,  Ar and Ca lines show small
discrepancies between the data  and the single temperature model.
These small  discrepancies sometimes give larger uncertainties in the strengths of 
faint lines such as K$\alpha$ lines of He-like Ar and Ca.
For example,  when fitting with a single temperature MEKAL model with the projected
spectrum at 2--4$'$,  temperature and
Ar abundance are determined as 2.2 keV and 0.7 solar, respectively.
However, when we restrict the energy range within 2.8--5.0 keV and fix
the  abundances of O, Ne, Mg, Si, S and Fe
to the best fit values obtained from the whole energy band fit, the 
temperature becomes  2.4 keV and Ar abundance  is determined as  1.1 solar.
Here, the  Ar abundance is changed by 60\%, although the best fit temperatures differ by only 10\%.
This small discrepancy between the S and Ar lines is also seen in the
deprojected spectra, although their errors are larger.

\begin{table*}[]
\caption{Ratios of K$\alpha$ lines and abundance ratios (errors correspond to
 68\%). Within 2$'$, the effect of the lower temperature component is
 considered.}
\begin{tabular}[t]{ccccccccccccc}
r   & H-O/H-Si & O/Si (solar)    & (H-S+He-S)/H-Si & S/Si (solar)
 &He-Ar/H-Si & Ar/Si (solar) & He-Ca/H-S & Ca/S (solar)\\ \hline
\multicolumn{9}{c}{Annular spectra of the EMOS}\\\hline                                                       
  0.0- 0.5   & $2.26\pm0.29$ & $0.43\pm0.06$ & $1.08\pm0.06$ & $1.14\pm0.06$  & $0.16\pm0.03$ & ---   & $0.20\pm0.04$ & ---\\ 
  0.5- 1.0   & $2.16\pm0.15$ & $0.48\pm0.03$ & $1.02\pm0.04$ & $1.05\pm0.05$  & $0.14\pm0.02$ & $0.79\pm0.11$   & $0.23\pm0.03$ & $1.29\pm0.16$ \\
  1.0- 2.0   & $1.90\pm0.11$ & $0.45\pm0.03$ & $0.94\pm0.03$ & $0.96\pm0.03$  & $0.18\pm0.03$ & $0.87\pm0.12$   & $0.28\pm0.03$ & $1.53\pm0.15$ \\
  2.0- 4.0   & $1.84\pm0.13$ & $0.46\pm0.03$ & $0.93\pm0.03$ & $0.97\pm0.03$  & $0.17\pm0.01$ & $0.78\pm0.06$   & $0.23\pm0.02$ & $1.16\pm0.10$ \\
  4.0- 8.0   & $2.38\pm0.10$ & $0.59\pm0.02$ & $0.82\pm0.03$ & $0.87\pm0.03$  & $0.14\pm0.02$ & $0.67\pm0.09$   & $0.24\pm0.02$ & $1.18\pm0.11$ \\
  8.0-13.5   & $2.45\pm0.20$ & $0.61\pm0.05$ & $0.62\pm0.05$ & $0.67\pm0.05$  & $0.14\pm0.03$ & $0.71\pm0.16$   & $0.32\pm0.06$ & $1.56\pm0.29$ \\\hline
\multicolumn{9}{c}{Deprojected spectra of the EMOS}\\\hline		        
  0.0- 0.5   & $2.08\pm0.42$ & $0.35\pm0.07$ & $1.09\pm0.17$ & $1.16\pm0.18$  & $0.18\pm0.06$ & ---   & $0.18\pm0.08$ & --- \\
  0.5- 2.0   & $2.06\pm0.23$ & $0.45\pm0.05$ & $0.96\pm0.04$ & $1.00\pm0.04$  & $0.16\pm0.02$ & $0.93\pm0.11$   & $0.29\pm0.04$ & $1.60\pm0.21$  \\
  2.0- 5.6   & $1.77\pm0.12$ & $0.44\pm0.03$ & $0.90\pm0.04$ & $0.93\pm0.04$  & $0.18\pm0.02$ & $0.85\pm0.10$   & $0.24\pm0.03$ & $1.24\pm0.16$ \\
  5.6-11.3   & $2.74\pm0.31$ & $0.68\pm0.08$ & $0.75\pm0.07$ & $0.80\pm0.08$  & $0.14\pm0.03$ & $0.67\pm0.15$   & $0.25\pm0.05$ & $1.26\pm0.26$ \\\hline
\multicolumn{9}{c}{Annular spectra of the EPN}\\\hline			        
  0.0- 0.5   & $2.00\pm0.33$ & $0.38\pm0.06$  & $1.11\pm0.11$ & $1.17\pm0.11 $& $0.30\pm0.05$ & ---  & $0.24\pm0.09$ & --- \\
  0.5- 1.0   & $1.87\pm0.20$ & $0.41\pm0.04$  & $1.04\pm0.06$ & $1.07\pm0.06 $& $0.15\pm0.03$ & $0.81\pm0.13$   & $0.24\pm0.06$ & $1.36\pm0.31$  \\
  1.0- 2.0   & $1.90\pm0.18$ & $0.44\pm0.07$  & $0.92\pm0.05$ & $0.94\pm0.05 $& $0.18\pm0.02$ & $0.88\pm0.10$   & $0.28\pm0.04$ & $1.50\pm0.22$  \\
  2.0- 4.0   & $1.87\pm0.23$ & $0.47\pm0.06$  & $0.93\pm0.05$ & $0.97\pm0.05 $& $0.17\pm0.02$ & $0.81\pm0.11$   & $0.35\pm0.05$ & $1.78\pm0.25$  \\
  4.0- 8.0   & $2.09\pm0.26$ & $0.52\pm0.06$  & $0.76\pm0.05$ & $0.82\pm0.05 $& $0.08\pm0.02$ & $0.41\pm0.09$   & $0.34\pm0.09$ & $1.69\pm0.44$  \\
  8.0-13.5   & $2.70\pm0.43$ & $0.67\pm0.11$  & $0.67\pm0.08$ & $0.73\pm0.09 $& $0.20\pm0.05$ & $1.02\pm0.23$   & $0.44\pm0.08$ & $2.14\pm0.39$  \\
\hline									      				       
\end{tabular} 								     
\end{table*}

Therefore, in this section, we determine strengths of emission lines,
 and then determine abundance ratios, considering the temperature dependence
 of line ratios.
We have fitted the projected and deprojected spectra within an energy
band around  a given line with   thermal bremsstrahlung  and gaussians.
Obtaining strengths of  K$\alpha$ lines of  He-like S and Ar,  
contributions of K$\beta$ lines of  H-like  Si and S were subtracted 
using the strengths of K$\alpha$ lines.
Here, we used the K$\beta$ to K$\alpha$ ratio of the MEKAL code.
Strengths of H-like K$\alpha$ lines of  O  were obtained from  spectra 
within the energy range of 0.55--1 keV fitted with a gaussian and two
APEC models with zero O abundance.
Those of Si were also obtained from fitting with gaussians and the two
temperature APEC model with zero Si abundance, using the 1.1--2.2 keV energy band.
We  selected line ratios whose temperature dependence is small,
 for  accurate measurements of abundance ratios.
The results are summarized in Table 3. 
\subsection{O vs. Si}

\begin{figure}[]
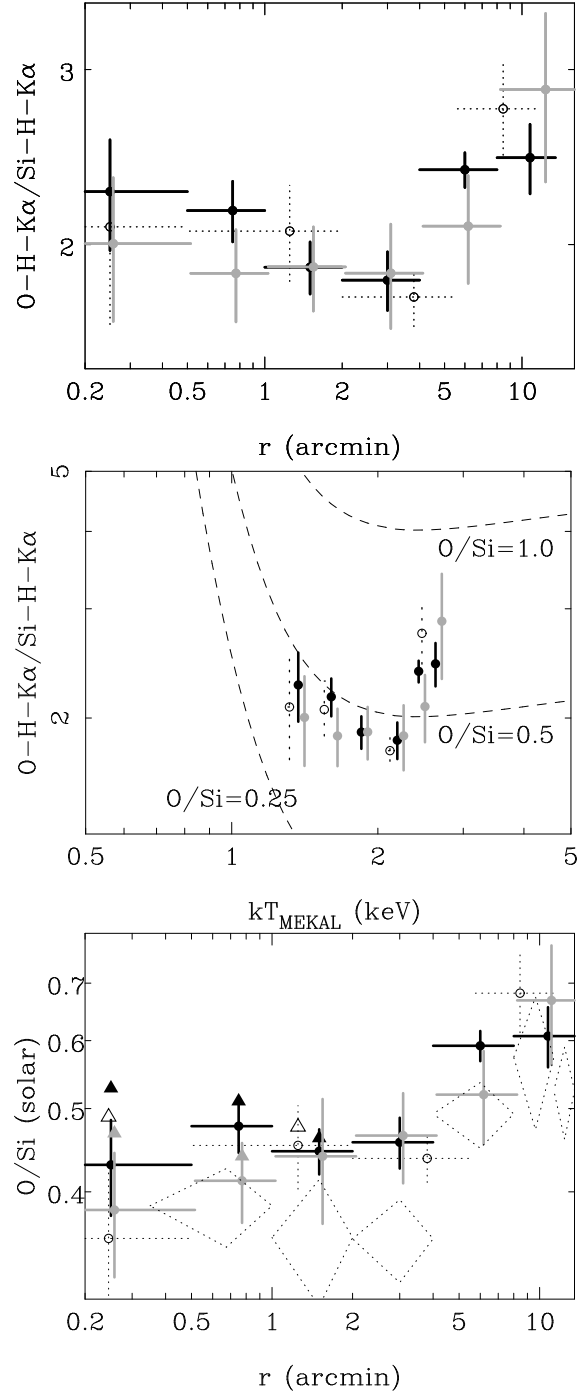

\begin{center}
  \resizebox{7.5cm}{!}{\includegraphics{matsushita_fig7a.ps}}
 \resizebox{7.5cm}{!}{\includegraphics{matsushita_fig7b.ps}}
 \resizebox{7.5cm}{!}{\includegraphics{matsushita_fig7c.ps}}
\end{center}
\caption{
The line ratios of K$\alpha$ lines of H-like O and Si of the EMOS
 (black) and the EPN (gray), plotted
 against the radius (upper panel) and the best fit temperature fitted
 with the single MEKAL model (middle panel).
Solid and dotted data correspond to projected and deprojected data, respectively.
Dashed lines in the middle panel correspond to constant abundance
 ratios in unit of the solar ratio.
The derived abundance ratios are plotted in the bottom panel.
Within 2 arcmin, contributions of the lower temperature component are
 considered. Triangles are the best fit values of the abundance ratio,
 assuming the temperature is larger than 1.7 keV.
The diamonds correspond to the abundance ratios derived from the deprojected  EMOS
 spectra  using the MEKAL model fit.
Errors represent the 68\% confidence level.
}
\label{osi}
\end{figure}

The abundance ratio of  O/Si is better obtained from the ratio of the line
strengths of K$\alpha$ lines of H-like ions.
Although the O and Si abundances  obtained from the EMOS and the EPN
using the same single temperature model  have  small discrepancies
of 20$\sim$ 30\%,  the  line ratios of the two detectors
 agree well with each other (Figure \ref{osi}).

The observed radial profile of the line ratio has a minimum at
$\sim$2$'$ (Figure \ref{osi}). 
Figure \ref{osi} also shows the line ratios plotted against the best
fit MEKAL temperatures from the whole energy band.
When the abundance ratio of O/Si is constant, the line ratio is  almost constant  above 1.7 keV and 
it starts to increase sharply below the lowest temperature.
Comparing the observed minimum value of $\sim$ 2 with the minimum value of $\sim4$
corresponding to the solar abundance ratio, no temperature distribution
can reproduce an  O/Si ratio larger than 0.5 solar at the radius of $\sim$2$'$.

Considering the temperature structure derived in Paper I, we can better
constrain the O/Si ratio.
The independence of the line ratio from the temperature above 1.7 keV
means that the abundance ratio can be directly
obtained  dividing the line ratio by that of the solar abundance
ratio when there is no temperature component below 1.7 keV.
Thus, beyond $2'$, the gradient of the line ratio reflects a change of the
O/Si ratio, since outside 2$'$, there is no temperature component below
1.7 keV (Paper I).
  From the line ratio at $r$= 2--4$'$, the O/Si is derived to be 0.46
solar,  and beyond $r>8'$, it increases to 0.6--0.7 solar (Figure \ref{osi}). 

Within 2$'$, the presence of the additional $\sim$ 1 keV temperature component
 must be considered.
The derived O/Si ratio  assuming that the temperature is larger than
1.7 keV should be the maximum value of the O/Si ratio which is $\sim$
0.5 solar within 2$'$ (triangles in the bottom panel of Figure \ref{osi}).
We also derive the O/Si ratio from the line ratio considering 
the fraction of the low temperature component, which is derived from the
spectral fit using the best fit two temperature APEC model.
Then,  the O/Si ratio is
derived to be 0.45 solar at 1--2$'$ and 0.4 solar within 0.5$'$.
Since the line ratio is a steep function when the temperature is around 1
keV, we have also checked a case when the lower temperatures is 
0.6 keV, which is 0.2 keV smaller than the central temperature.
For the innermost region, the derived O/Si ratio with the temperature of
 0.6 keV is consistent within 10 percent,
since the temperature of the softer component
becomes lower, the fraction of the component also becomes lower which is
derived from the spectral fitting from Fe-L.

 For comparison, we also plotted the O/Si ratio
derived from the deprojected spectra through  the spectral fits and the
line ratio. 
Although the O/Si ratios derived from the line ratios are systematically
larger by $\sim$ 20\% than those derived from spectral fit with
the MEKAL model, the gradient of the O/Si ratio is consistent between
the two, and the O/Si ratio changes by a factor of 1.4 from the center to 10$'$.

\begin{figure}[]
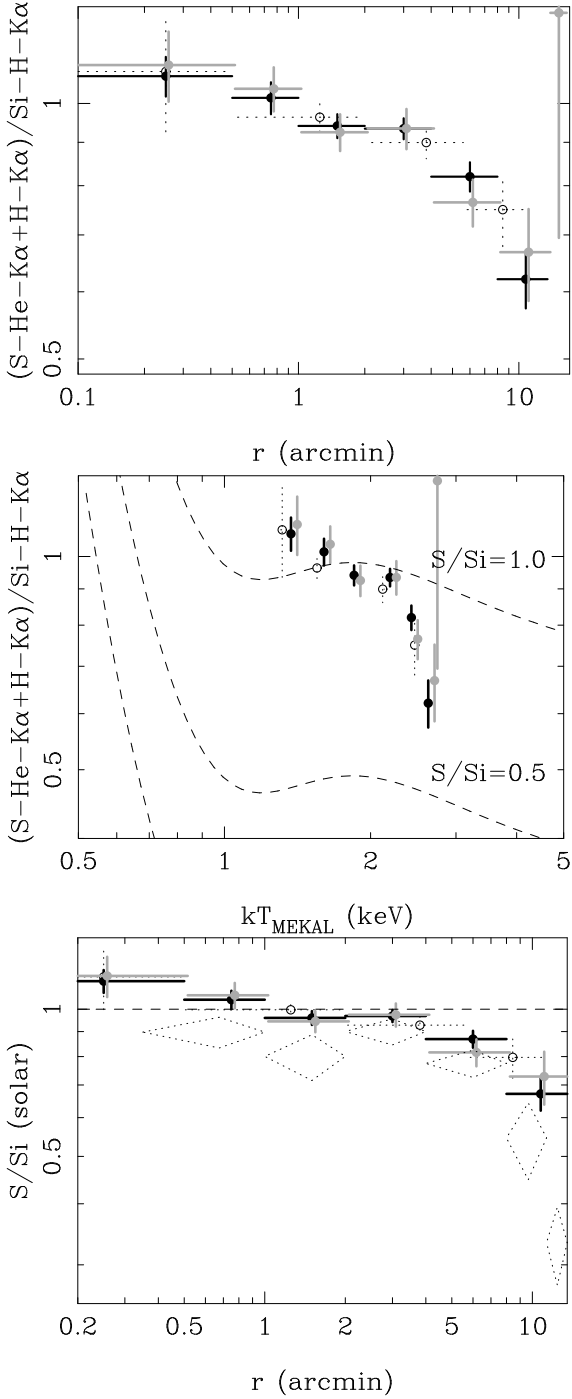

\begin{center}
 
 \resizebox{7.5cm}{!}{\includegraphics{matsushita_fig8a.ps}}
 \resizebox{7.5cm}{!}{\includegraphics{matsushita_fig8b.ps}}
 \resizebox{7.5cm}{!}{\includegraphics{matsushita_fig8c.ps}}
\end{center}
\caption{The line ratios of K$\alpha$ lines of the sum of He-like and H-like S to H-like Si of the EMOS
 (black) and the EPN (gray), plotted against radius (top panel) 
and temperature (middle panel), and the radial profile of the S/Si ratio
 (bottom panel). The meaning of the symbols are the same as
 in Figure 7. Errors represent the 68\% confidence level.}
\label{ssi}
\end{figure}

\subsection{S vs. Si}

Although the ratios of K$\alpha$ lines of He-like or H-like S to that of H-like
Si are both steep functions of temperature,
the ratio of the sum of  the two K$\alpha$ lines of S to the 
H-like Si line is  nearly constant within 10\% between 0.9 to 5 keV when
the S/Si ratio is constant (Figure \ref{ssi}).
The important point is that we can derive the abundance ratio almost
independently from the temperature structure of the ICM.
Considering that there is no temperature component below 1 keV, 
 as in the case of the O/Si ratio, even for the projected data, the S/Si ratio can be
directly calculated  from the line ratio.
 As in the previous subsection, the effect of the lower temperature
 component is estimated to be less than a few percent.

The observed  profile of the line ratio shows a negative gradient and 
at $r$=10$'$,   it is about  half of the central value.
This gradient reflects the change of the S/Si ratio, because of the
small dependence on temperature.
Converted to the abundance ratio, the S/Si ratio
is 1.0 solar within 2$'$, 
and drops to 0.67 solar at $\sim$ 10$'$ (Figure \ref{ssi}).
 These results are systematically larger than those derived from the
spectral fits.
Especially, at $r>8'$, the spectral fits on the projected spectra give
the S/Si ratio to be 0.3--0.5 solar (Table 1, Figure 2), 
while the line ratio gives the value of 0.6--0.7 solar.
For the projected data, the S abundance from the spectral fit should not
be correct since the single temperature model cannot fit the two K$\alpha$
lines simultaneously as shown in Figure 6.

\subsection{Ar vs. Si}

\begin{figure}[]
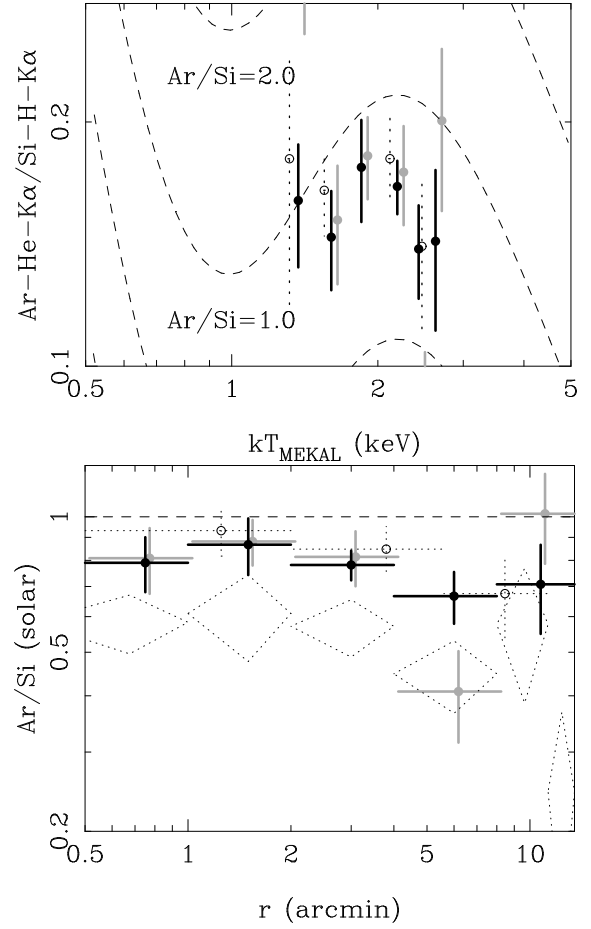

\begin{center}
 \resizebox{7.5cm}{!}{\includegraphics{matsushita_fig9a.ps}}
 \resizebox{7.5cm}{!}{\includegraphics{matsushita_fig9b.ps}}
\end{center}
\caption{The line ratios of K$\alpha$ lines of He-like Ar to H-like Si of the EMOS
 (black) and the EPN (gray) (top panel) and the Ar/Si ratio (bottom
 panel).
The meaning of the symbols are the same
as in Figure 7. Errors represent the 68\% confidence level.}
\label{arsi}
\end{figure}

Although the line ratio of the K$\alpha$ lines of He-like Ar to H-like
Si is a steeper function of temperature than S/Si or O/Si ratio,  
within 1.7 to 2.8 keV,  
at a given ratio of Ar/Si, the change of the line ratio  is less than 10 \% (Figure \ref{arsi}).
In Paper I,  we found that at $R>0.5'$, the ICM temperature at a given radius is 
dominated by a single temperature component whose temperature is larger
than 1.7 keV.  
From Figure 7 in  Paper I,
 the contributions from the  1 keV component to Ar and H-like Si lines
 are negligible.
Therefore,  as in case of the S to Si abundance ratio, 
we can obtain the  abundance ratio of Ar/Si from the line ratio.
 The results indicate that the Ar/Si is 0.7--0.8 solar. 

The derived Ar/Si ratios are systematically larger than the results of
the spectral fit on the projected data  (Finoguenov et al. 2002) by 30\% within 3$'$ and by a
factor of 2 at $r>8'$.  This large discrepancy is caused by the 
failure to fit the continuum between the S and Ar lines illustrated in
Figure 6.

\subsection{Ca vs. S }

\begin{figure}[]
\begin{center}
 
 \resizebox{7.5cm}{!}{\includegraphics{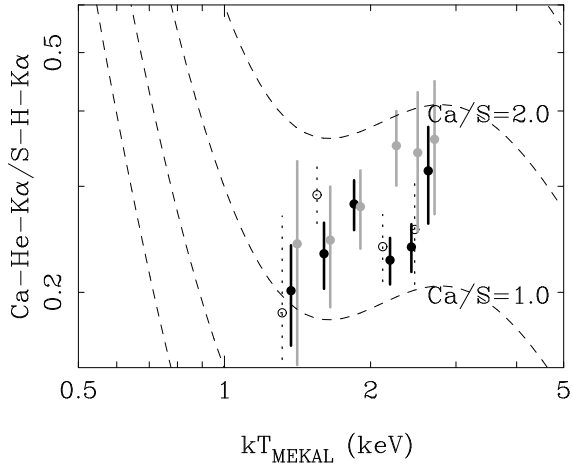}}
\end{center}
\caption{The line ratios of K$\alpha$ lines of He-like  Ca to H-like S of the EMOS
 (black) and the EPN (gray). The meaning of the symbols are the same
 as in Figure 7. Errors represent the 68\% confidence level.}
\label{cas}
\end{figure}

The  ratio of the K$\alpha$ line of the He-like Ca to that of the
H-like S also show a small temperature dependency (Figure \ref{cas}). 
Because these lines are dominated by a hotter temperature component
even around the center,  we can calculate the Ca/S ratio in the same way.
The Ca/S ratio is $\sim$ 1.5 solar within the whole region.

\subsection{Mg vs. O}

\begin{figure}[]
\resizebox{8cm}{!}{\includegraphics{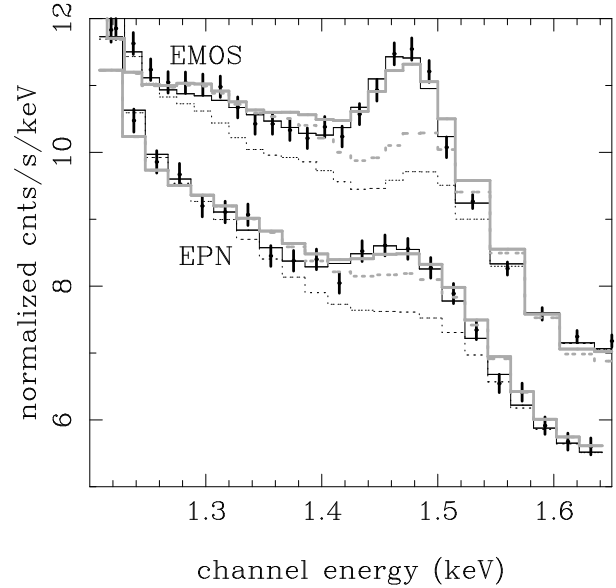}}
\caption{The projected spectra of $r$=2-4$'$ at the energy band of
 Mg of the EMOS  and the EPN fitted with the two temperature APEC model with zero Mg
 abundance (black dotted lines) and gaussians.
Those using the MEKAL model are shown as gray lines.
}
\label{mgkspec}
\end{figure}

The energy of the K$\alpha$ line of  H-like Mg  is slightly shifted from
the peak of the Fe-L emission at $\sim$ 1.49 keV (Figure \ref{mgkspec}).
Thus, the K$\alpha$ line can be distinguished within the energy
resolution of the CCDs.
The problem with the spectral fitting is the modeling of the
Fe-L structure around the Mg line.
Due to the problem of the strong instrumental line of the EMOS,
 we do not present the results on the Mg-K line for $r>8'$.

The strengths of H-like and He-like K$\alpha$ lines of Mg  were obtained from spectra
within 1.1--1.6 keV, fitted with  gaussians and  a two  temperature
MEKAL or  APEC model with zero Mg abundance.
The results are summarized in Table 4.

\begin{figure}[]
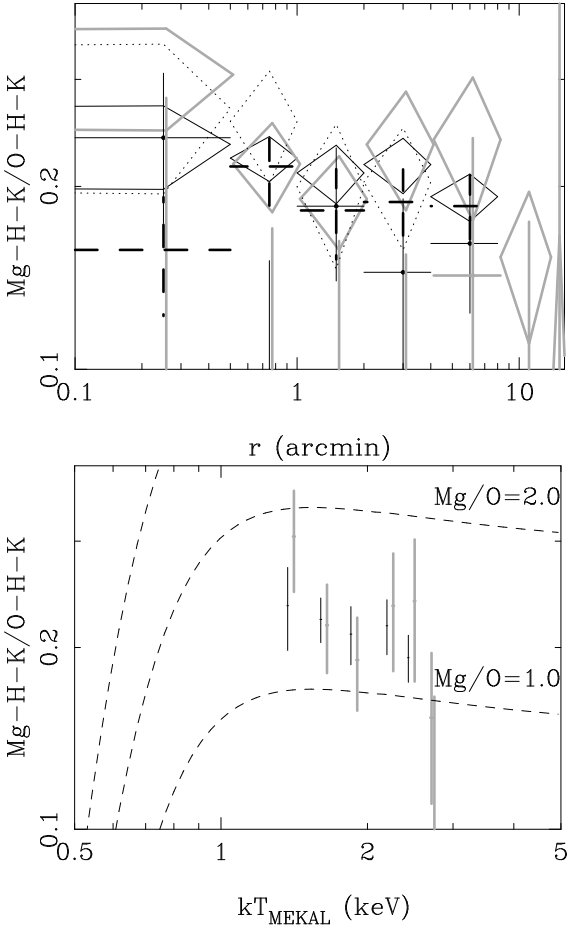

\resizebox{7.5cm}{!}{\includegraphics{matsushita_fig12a.ps}}
\\
\resizebox{7.5cm}{!}{\includegraphics{matsushita_fig12b.ps}}
\caption{
(Top panel) The ratio of strength of K$\alpha$ lines of H-like Mg and O of
 the projected spectra of the EMOS (black) and EPN (gray).
The Fe-L is modeled with the APEC (diamonds) and the MEKAL (crosses)
 and bremsstrahlung and gaussians (bold dashed lines).
The results of the deprojected spectra of the EMOS, where the Fe-L lines are
 modeled with the APEC are shown as dotted diamonds.
(Bottom panel) The line ratios when the Fe-L  lines are modeled with the APEC are plotted against the best fit temperature
 fitted with the single MEKAL model. 
Dashed lines correspond to  constant abundance ratios.
Errors represent the 68\% confidence level.
}
\label{mgo}
\end{figure}

Figure \ref{mgo} shows the ratios of the K$\alpha$ line of the
H-like Mg to that of H-like O.
The strengths of the Mg line obtained  using the APEC Fe-L model are 
$\sim$ 30\% larger than those derived using the MEKAL model.
Since the APEC fit to the Fe-L/Mg-K feature of 1.4 keV is  better
than the MEKAL model fit,  the APEC code may be better to describe the
Fe-L feature,
although the discrepancy between the data and model of the MEKAL plus gaussians is only
several percent.

\begin{table}
\caption{Line ratio of K$\alpha$ of H-like Mg, O, and Si and their abundance
 ratios, when the Fe-L is modeled with the APEC (errors correspond to 68\%)}
\begin{center}
  \begin{tabular}[t]{cccccccc}
r   & H-Mg/H-O & Mg/O {\small (solar)}  & H-Mg/H-Si & Mg/Si {\small (solar)} \\ \hline 
\multicolumn{5}{c}{Annular spectra of the EMOS}\\\hline  
  0.0- 0.5   & $0.23\pm0.04$ & $1.46\pm0.23$   & $0.53\pm0.05$ & $0.63\pm0.07$ \\
  0.5- 1.0   & $0.22\pm0.02$ & $1.33\pm0.11$   & $0.48\pm0.03$ & $0.64\pm0.04$ \\
  1.0- 2.0   & $0.21\pm0.02$ & $1.26\pm0.14$   & $0.40\pm0.04$ & $0.57\pm0.06$ \\
  2.0- 4.0   & $0.22\pm0.02$ & $1.30\pm0.14$   & $0.40\pm0.03$ & $0.59\pm0.05$ \\
  4.0- 8.0   & $0.19\pm0.02$ & $1.16\pm0.10$   & $0.46\pm0.04$ & $0.69\pm0.06$ \\\hline
\multicolumn{5}{c}{Deprojected spectra of the EMOS}\\\hline  
  0.0- 0.5   & $0.27\pm0.07$ & $1.73\pm0.48$  & $0.56\pm0.13$ & $0.61\pm0.14$ \\
  0.5- 2.0   & $0.20\pm0.03$ & $1.19\pm0.17$  & $0.41\pm0.04$ & $0.54\pm0.05$ \\
  2.0- 4.0   & $0.20\pm0.05$ & $1.21\pm0.28$  & $0.42\pm0.08$ & $0.66\pm0.10$ \\\hline
\multicolumn{5}{c}{Annular spectra of the EPN}\\\hline  
  0.0- 0.5   & $0.31\pm0.06$ & $1.90\pm0.36$  & $0.61\pm0.08$ & $0.72\pm0.10$ \\
  0.5- 1.0   & $0.22\pm0.04$ & $1.31\pm0.22$  & $0.41\pm0.06$ & $0.54\pm0.07$ \\
  1.0- 2.0   & $0.19\pm0.03$ & $1.14\pm0.20$  & $0.36\pm0.06$ & $0.52\pm0.08$ \\
  2.0- 4.0   & $0.23\pm0.05$ & $1.41\pm0.31$  & $0.44\pm0.08$ & $0.65\pm0.13$ \\
  4.0- 8.0   & $0.24\pm0.06$ & $1.45\pm0.38$  & $0.50\pm0.12$ & $0.75\pm0.18$ \\
  8.0-13.5   & $0.15\pm0.04$ & $0.94\pm0.26$  & $0.41\pm0.10$ & $0.63\pm0.15$ \\\hline
 \end{tabular}
\end{center}
\end{table}

\begin{figure}
  \resizebox{7.5cm}{!}{\includegraphics{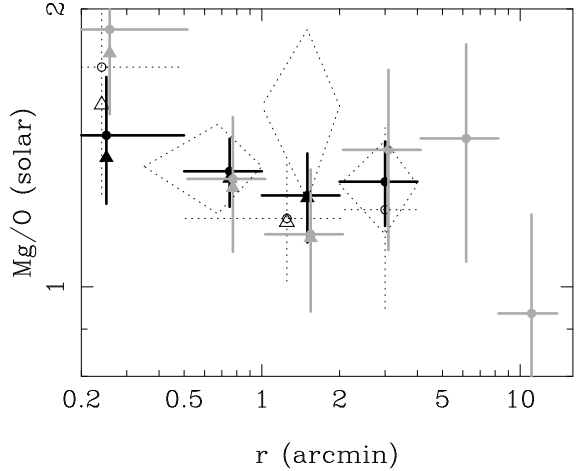}}
\caption{ The Mg/O ratio  derived from the line ratio, where the
 Fe-L lines are modeled with the APEC code.
 The meaning of the symbols are the same  as in Figure 7.
The dotted diamonds correspond to the abundance ratios of the EMOS derived from
 the APEC model fit to the deprojected spectra.
Errors represent the 68\% confidence level.}
\label{mgoa}
\end{figure}

The projected EMOS spectra in the energy range of 1.35 to 1.6 keV are also
fitted with a bremsstrahlung model plus two gaussians at 1.47 keV and 1.49
keV, since the K$\alpha$ line of H-like Mg at 1.47 keV  and a Fe-L peak at 1.49 can be
distinguished within the EMOS energy resolution (Figure \ref{mgkspec}).
Freeing the strength of 1.49 keV peak of the Fe-L enable us to constrain
 the Mg line strength.
The derived strengths of the  Mg line are consistent with those derived using the APEC
code for the modeling of the Fe-L.
This result supports the Mg line strengths from the APEC code than those
from the MEKAL code.

As in Sec. 5.1-5.4, we have plotted the line ratio from the APEC code against the best
fit MEKAL temperature (Figure \ref{mgo}).
The constant Mg/O ratio gives a constant line ratio within 10\% above 1 keV.
 As in Sec. 5.1, we have estimated the effect of the temperature
component below 1 keV, which is less than 10\% for the innermost region.
The derived Mg/O ratios are about 1.2--1.3 solar with no radial
gradient at $r>0.5'$, when we adopt the Fe-L modeling for the APEC
(Figure 13).
Within 0.5$'$, a higher Mg/O ratio is allowed,
since adding a temperature component blow 1 keV, the Mg/O ratio
increases.

The O and Mg abundances obtained from the RGS spectrum through a
spectral fit are
0.49$\pm0.04$ solar and 0.9$\pm0.2$ solar, respectively (Sakelliou et al
2002).
This Mg/O ratio derived from the RGS is a factor of 1.5 larger than
that at $r>0.5'$ derived from the EMOS spectra using the APEC code for modeling
the Fe-L.
Since most of the photons detected by the RGS come from the very center,
where a larger Mg/O ratio is also allowed by the EMOS data, 
the Mg/O ratio may increase at the center.

\subsection{Mg vs. Si}

\begin{figure}[]
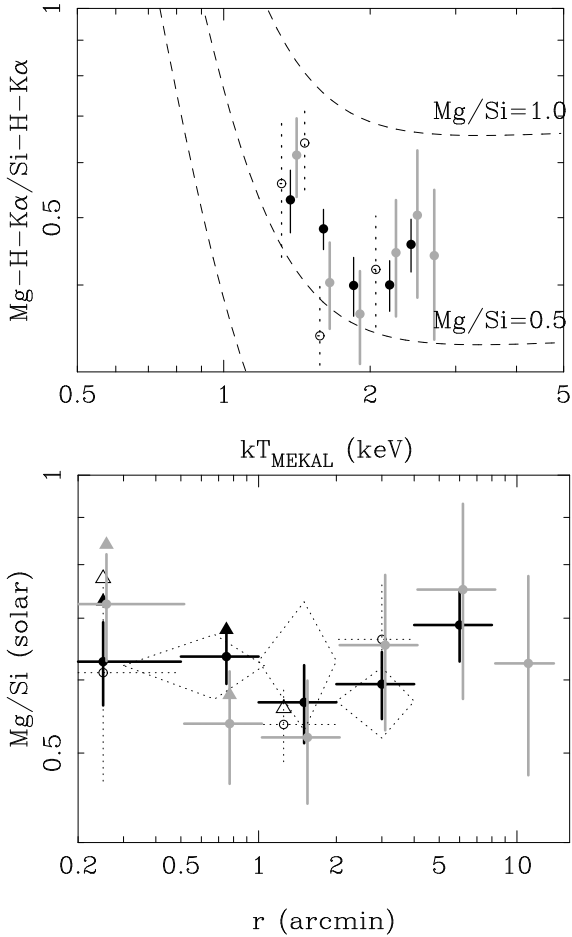

\begin{center}
 \resizebox{7.5cm}{!}{\includegraphics{matsushita_fig14a.ps}}
 \resizebox{7.5cm}{!}{\includegraphics{matsushita_fig14b.ps}}
\end{center}
\caption{
(Top panel)  The ratio of strength of K$\alpha$ lines of H-like Mg and Si 
plotted against the best fit temperature fitted with the single MEKAL model.
The Fe-L is modeled with the APEC.
Dashed lines correspond to  constant abundance ratios.
The meanings of the symbols are the same as in Figure 7.
(Bottom panel) The radial profile of the Mg/Si ratio derived from the line ratio.
The dotted diamonds are the Mg/Si ratios derived from the spectral fitting using the APEC code.
Errors represent the 68\% confidence level.
}
\label{mgsi}
\end{figure}

The derived ratios of the K$\alpha$ lines of H-like Mg and Si are
summarized in  Table 4 and Figure \ref{mgsi}.
Here, only the result using the APEC code for the Fe-L modeling is
shown.
The temperature dependence of the line ratio is quite similar to that of
the K$\alpha$ lines of O and Si, and the Mg/Si ratio was derived in the
same way as in Sec. 5.1.
Within 2$'$, the contribution of the 1 keV component is taken into account.
The derived Mg/Si ratio is $\sim 0.6$ solar, which agrees well
with the ratio derived from the spectral fit using the APEC model in Sec
4.

\subsection{Fe vs. Si}
\begin{figure}[]
\begin{center}
  \resizebox{7.5cm}{!}{\includegraphics{matsushita_fig15a.ps}}
 \resizebox{7.5cm}{!}{\includegraphics{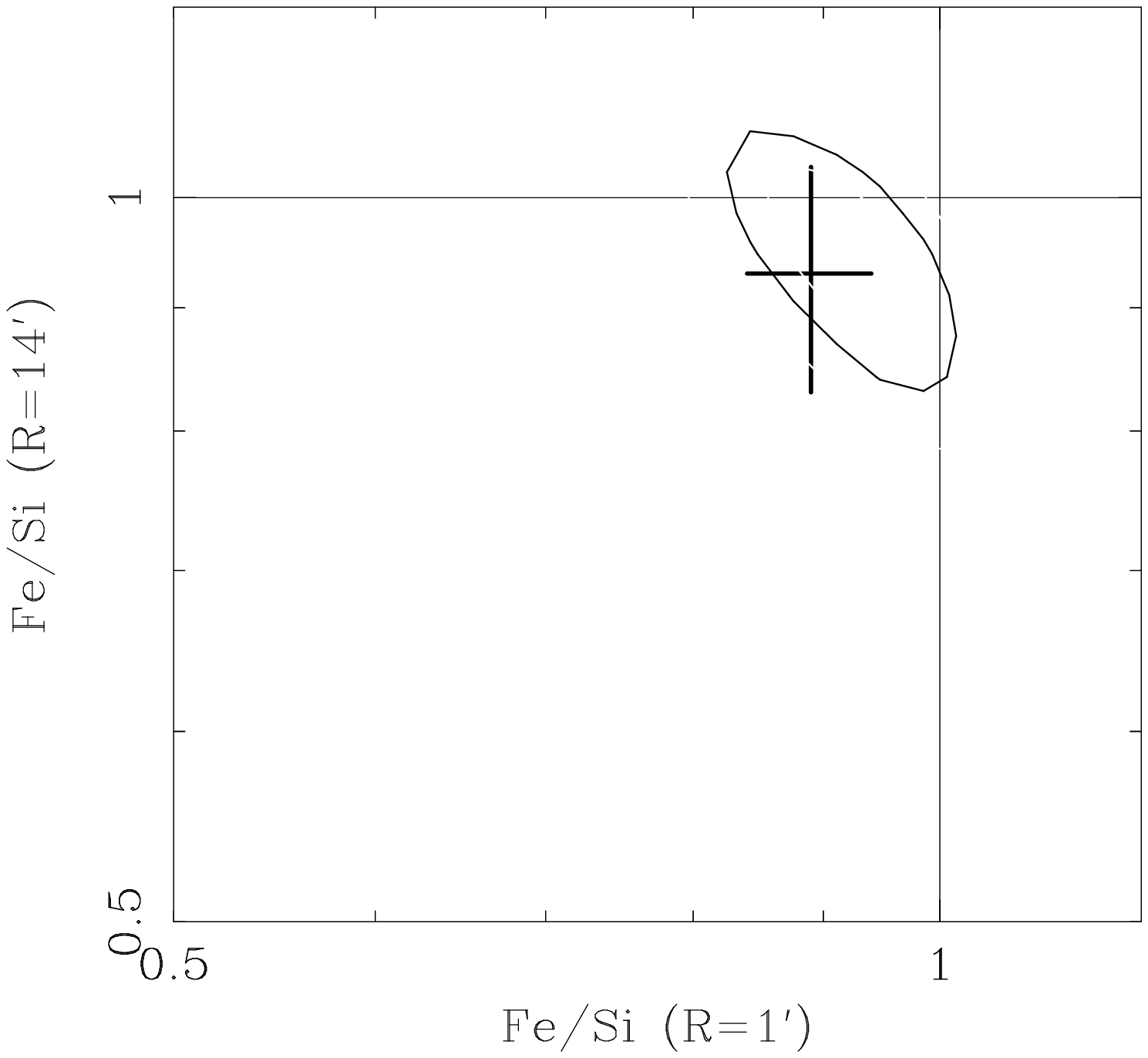}}
\end{center}

\caption{
 (upper panel) The radial profile of projected line 
 ratios of the EMOS (black) and EPN (gray) between K$\alpha$ of
 H-like Si and He-like Fe. Errors correspond to 68\% confidence level.
The dashed line represents the best fit model using the temperature
 profile obtained in Paper I.
(lower panel) 90 \% confidence contour of the Fe/Si at $R$=14$'$ vs. that at
 $R$=1$'$ obtained from the ratio of the K$\alpha$ line of the He+Li-like Fe to
 H-like Si of the projected data, assuming Fe/Si=$a+b R$ and kT=$1.68 (1+(R/1.6')^2)^{0.115}$ (keV). 
 The cross is the correlation derived from the
 deprojected spectra,  determined from the Fe-L emission.
 }
\label{fesiratio}
\end{figure}

In contrast to the temperature dependences of the line ratios in the
previous subsections,
the ratio of the Fe-K line at 6.7 keV (which includes He and Li like ions)
to the K$\alpha$ line of the H-like Si line strongly depends on
temperature.
For example, at 2 keV, a 10 \% increment of temperature increases the
Fe/Si ratio by 36 \%.
As a result, small uncertainties in the temperature structure give large
uncertainties in the abundance ratios.
Therefore, the Si/Fe ratio cannot obtained from the line ratio as in
previous subsections. 
Since this ratio is quite important, 
we have calculated the Si/Fe line ratio of the projected data
using the temperature structure obtained in Paper I and compared it with
the observed profile  (Figure \ref{fesiratio}).
We used the relation of $kT=1.68 (1+(R/1.6')^2)^{0.115}$ keV, since the
hotter component dominates these lines even within 2$'$.
Assuming the Fe/Si ratio is approximated by $a+bR$, where a and b are
free parameters, we fitted the profile of the line ratio.
The radial profile of the line ratio is well fitted with the model
($\chi^2$=7.67 for 9 degrees of freedom)
and the Fe/Si ratio is determined to be $\sim$ 0.9 solar within the whole
field of view (Figure \ref{fesiratio}),
although the abundance ratio changes by 20\% when the whole temperature is shifted by 5\%.
Considering that the value of 0.9 solar is quite close to the
ratio derived from the spectral fitting of the deprojected data, where
the Fe abundance is determined by the Fe-L emission,
the Fe/Si ratio should be $\sim$ 0.9 solar and its radial gradient
is  less than 10--20\%.

\section{Effect of resonance line scattering}

Some resonant lines may become optically thick in the dense cores of
clusters.   Shigeyama (1998) calculated the effect on M~87 and several 
X-ray luminous elliptical galaxies, and found that it should be important at
the center.  B\"ohringer et al. (2001) suggested that the observed
central abundance drop may be due to the effect.
In the case of an  X-ray luminous elliptical, NGC 4636, Xu et al. (2002)
discovered direct evidence of resonant scattering using a line ratio
between optically thick and thin lines.

The profile of optical depth for resonant lines of some prominent emission
lines  are calculated in B\"ohringer et al. (2001). 
Using the observed abundance profiles (Table 1), the temperature profile
 used in Sec. 5.7
and the density profile derived in Paper I,  optical depths from the center
 for K$\alpha$ lines of H-like Si and O are calculated to be 1 and 0.4,
 respectively.
Therefore,  we must evaluate the effect of the  scattering,
although  the central abundance drop is very small when fitted with the two temperature MEKAL model.

\begin{figure}[]
\resizebox{\hsize}{!}{\includegraphics{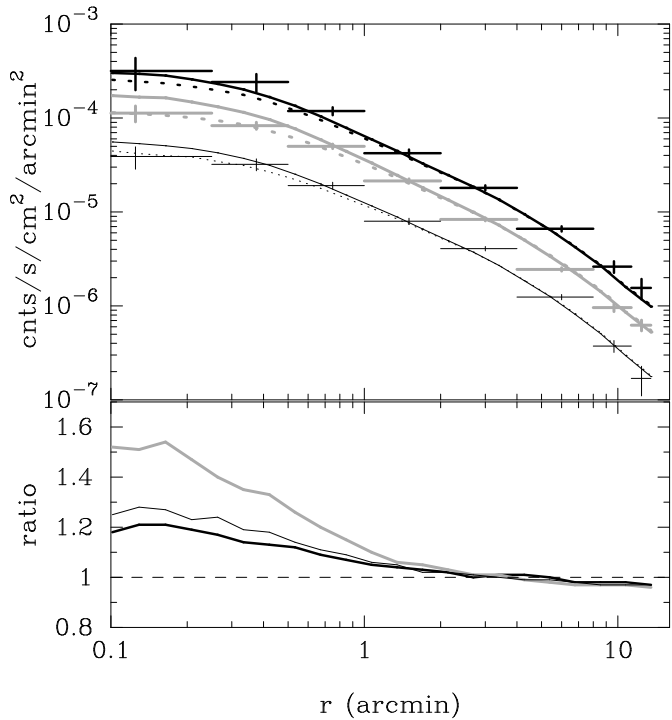}}
\caption{
Radial profile of brightness  of H-like ions of O (bold black lines), Si
 (gray lines), and S (thin black lines).
 Crosses are observed profiles and solid and dotted
 lines corresponds to the simulated profiles before and after resonant line
 scattering, respectively. The bottom panel shows the ratios of the
 profiles of before and after the scattering.}
\label{res}
\end{figure}

Ignoring turbulent motion, we have calculated the effect of resonant
line scattering on the K$\alpha$ lines of H-like ions, using
 Monte-Carlo simulations.

The results are summarized in Figure \ref{res}.
The scattered profiles of lines are consistent with those
ignoring scattering beyond 1$'$.
Within this radius, the brightness of the lines decrease.
The reduction rates at 0.5$'$ are 10, 30, 15\% for O, Si, and S, respectively.
The effect is important only within 0.5$'$ on the Si line,
where the uncertainty in temperature structure also gives an
uncertainty in abundance determination.
\section{Discussion}

\subsection{Summary of abundances}

The abundance profiles of O, Si, S, Ar, Ca and Fe are obtained from the deprojected
spectra, based on the temperature study in Paper I.
Although  version 1.0 of the APEC code gives significantly different
results (Paper I),  version 1.1 of the APEC code gives consistent
results with those of the MEKAL code for these elements.
Within 2$'$, considering the two temperature nature of the ICM,
the abundance profiles from deprojected spectra become almost constant.
Within this radius, the Si, S, Ar, Ca and Fe abundances are $\sim$ 1.5 solar, while the 
 O  abundance is a factor of 2 smaller.
 Using the APEC code, the Mg abundance is derived to be $\sim$ 1 solar at
the center, since the APEC code gives a better fit on the Fe-L/Mg-K structure at 1.4 keV.
Beyond 2$'$,  Si, S, Ar, Ca and Fe show strong negative gradients, while
the abundance gradient of O is smaller.
The Si/Fe ratio is 1.1 solar, with no gradient in the field of view.

The projected and deprojected profiles of abundance ratios are also obtained through 
line ratios.
Since  some line ratios among $\alpha$-elements are nearly constant within
the temperature range of the ICM around M~87,  the abundance ratios are
obtained with better accuracy.
From the temperature dependence of the line ratio, we can conclude that 
the low O/Si ratio cannot be explained by any temperature model, and
the ratios of O/Si are  0.4--0.5 solar at the center and 0.6--0.7 solar at the
outer region.
The Mg/O ratio is determined to be 1.2--1.3 solar.
The Ar and S abundances thus obtained are systematically larger than the
spectral fit results due to small disagreements in the continuum fits around
these lines.
The S/Si is  1.1 solar at the center and  0.7 solar at outer regions.
The Si/Fe ratio derived from the projected profile of the
K$\alpha$ line ratio is consistent with that derived from the
deprojected spectra through the spectral fit, where the Fe abundance is
determined by the Fe-L structure.

The observed abundance pattern is similar to those obtained for A 496
(Tamura et al. 2001), and NGC 4636 (Xu et al. 2002). 
Therefore, this abundance pattern may be uniform around the central
galaxies of groups or clusters.

\subsection{Comparison of the abundance pattern with Galactic stars}

\begin{figure}[]
\resizebox{\hsize}{!}{\includegraphics{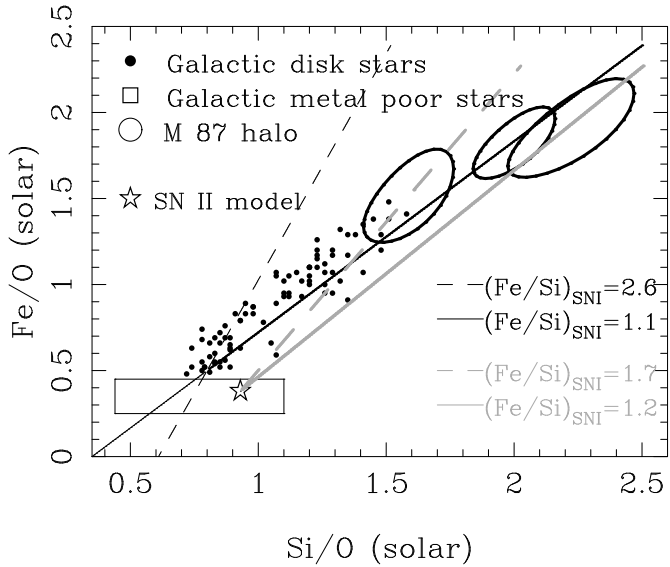}}
\caption{The Fe/O ratio is plotted against the Si/O ratio.
Those of disk stars (Edvardsson et al. 1993; closed circles) and
 average values of metal poor stars
 (Clementini et al. 1999; open square) in the Galaxy   are also plotted.
  The black solid line and
 dashed line represent the relation of the abundance pattern with
 (Fe/Si)$_{\rm SN Ia}$ =1.1 solar (best fit relation) and  (Fe/Si)$_{\rm SN Ia}$=2.6  solar (W7 ratio; Nomoto et al. 1984), respectively.
  The gray solid line and dashed line represent the relation of 
  (Fe/Si)$_{\rm SN Ia}$=1.2 and 1.7 solar, respectively, adopting the
 abundance pattern of the SN II by Iwamoto et al (1999; asterisk) 
using the nucleosynthesis model of Nomoto et al. (1997) and assuming a Salpeter IMF.
}
\label{stars}
\end{figure}

\begin{figure}[]
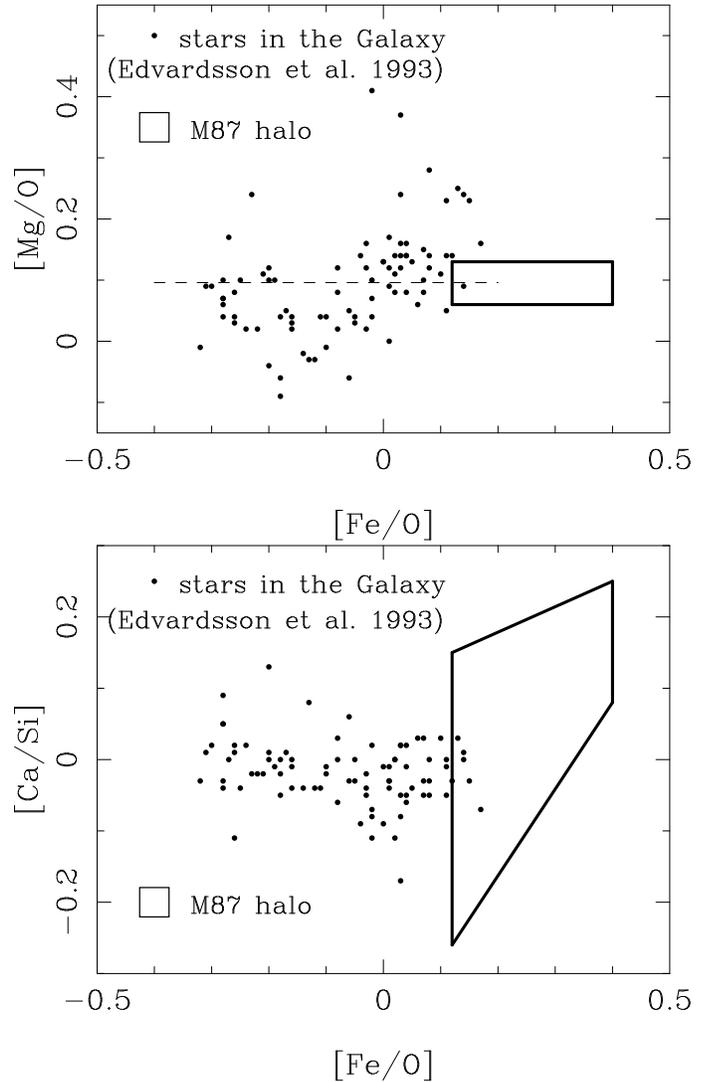

\resizebox{9cm}{!}{\includegraphics{matsushita_fig18a.ps}}
\resizebox{9cm}{!}{\includegraphics{matsushita_fig18b.ps}}
\caption{
[Mg/O], and [Ca/Si] are plotted against [Fe/O]. 
Those of disk stars (Edvardsson et al. 1993; closed circles) 
 in the Galaxy   are also plotted.
}
\label{stars2}
\end{figure}

Figures \ref{stars} and \ref{stars2} show the observed abundance ratios of O, Mg, Si, Ca
and Fe compared to those of the Galactic disk stars (Edvardsson et
al. 1993).
The average abundance ratios of  low metallicity (i.e. [Fe/H]$<$-0.8)
stars of the Galaxy by Clementini et al. (1999), which should reflect
the average abundance pattern of SN II products of our Galaxy are also
plotted.
Other papers on Galactic low metallicity stars give similar abundance
ratios. [O/Fe] ratios for stars with [Fe/H]$<-1$ derived by Nissen et al. (1994) and Peimbert (1992) are
0.48$\pm0.16$, and 0.5, respectively.
Gratton \& Sneden (1991) gives [Si/Fe] of 0.3 for these stars.

The observed abundance pattern of M~87 is located at a simple extension of that of  Galactic stars,
although the observed Fe/O range of M~87 is systematically larger.

The Mg/O ratio of the ICM is plotted assuming the ratio is constant
within the observed region.
We adopted the ratio derived from the line ratio, using the Fe-L modeling with the APEC code, 
since  APEC gives better fits around the Mg-K/Fe-L region at 1.3--1.5
keV.
[Mg/O] of the Galactic stars is almost constant at the same value 
as the ICM around M~87. 
The Galactic [Ca/Si] tends to  a slightly smaller value than the that of M~87,
but the difference is only 0.1 dex.

We note that Xu et al. (2002) also discovered similar values of the
Fe/O and Mg/O ratio  for the ISM in an elliptical galaxy, NGC 4636.
The Si/Fe ratio of this galaxy observed by ASCA is also determined to be
$\sim$ 1 solar (Matsushita et al. 1997; 2000).
Thus, the O/Mg/Si/Fe pattern of NGC 4636 is similar to that of M~87. 
Therefore, the abundance pattern of the ICM of M~87 is not peculiar, but
consistent, not only with  an elliptical galaxy  NGC 4636, but also with that of our Galaxy.

\subsection{Nucleosynthesis from SN Ia}

The observed radial change of the O/Fe ratio indicates  that 
the contribution from SN Ia increases toward center.

Although  the abundance pattern of ejecta of SN II may differ between
early-type and late-type galaxies, and that of SN Ia also may not be a
constant (Umeda et al. 1999; Finoguenov et al. 2002),  for a first
attempt we have assumed  (Fe/Si)$_{\rm SN Ia}$,
 (Si/O)$_{\rm SN II}$, and  (Fe/O)$_{\rm SN II}$ to be constants.
Here,  (Fe/Si)$_{\rm SN Ia}$ is the Fe/Si ratio of ejecta of SN Ia, and
(Si/O)$_{\rm SN II}$ and (Fe/O)$_{\rm SN II}$  are the Si/O ratio and
the Fe/O ratio of the ejecta of SN II, respectively.
Since O is not synthesized by SN Ia, the Fe/O ratio is then expressed by,
\begin{equation}
  \frac{\rm{Fe}}{\rm{O}} = \left(\rm{\frac{Fe}{Si}}\right)_{\rm{SN Ia}}\left( \left(\frac{\rm{Si}}{\rm{O}}\right)-\left(\rm{\frac{Si}{O}}\right)_{\rm{SN II}}\right)+\left(\rm{\frac{Fe}{O}}\right)_{\rm{SN II}}
\end{equation}
Thus, (Fe/Si)$_{\rm SN Ia}$  is the inclination in the Si/O vs. Fe/O
 plot (Figure \ref{stars}).
Even within the  M~87 data, the gradient indicates that (Fe/Si)$_{\rm SN Ia}$
at the center is smaller than $\sim$ 1.3 solar.
When we adopt the abundance pattern of the Galactic metal poor stars by
Clementini et al. (1999) as that of SN II,  (Fe/Si)$_{\rm SN Ia}$ around M~87 is
determined to be $\sim 1.1 $ solar  (Table \ref{sn1} ).
In addition, the M~87 data are located at the extension of those of
Galactic stars, and the whole trend is consistent with (Fe/Si)$_{\rm SN
Ia}$ to be $\sim$ 1.1 solar.
This means that also for metal rich stars in the Galaxy, the (Fe/Si)$_{\rm SNIa}$  is determined to
be $\sim$ 1.3 solar.
These values are a factor of 2 smaller than the standard W7 model (Nomoto et
al. 1984; Table \ref{sn1}), 
and closer to the WDD1 ratio (Iwamoto et
al. 1999), which considers slow deflagration.

\begin{figure}[]
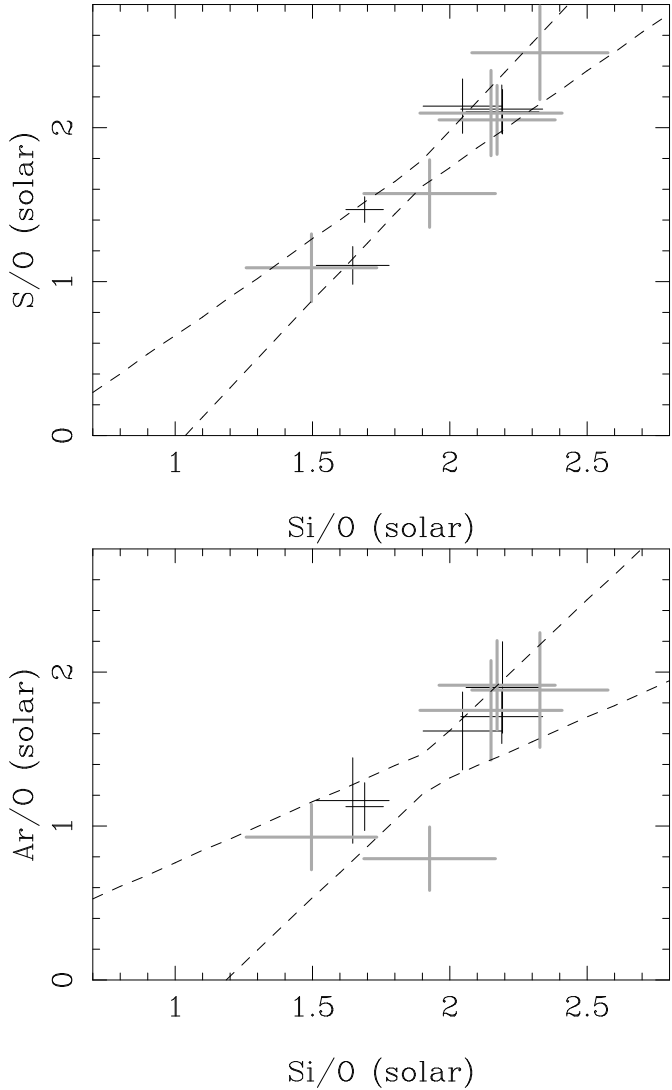

\resizebox{\hsize}{!}{\includegraphics{matsushita_fig19a.ps}}
\resizebox{\hsize}{!}{\includegraphics{matsushita_fig19b.ps}}
\caption{
The S/O and Ar/O ratio are plotted against the Si/O ratio observed by
 the EMOS (black) and the EPN (gray).  Dashed lines represents the
 average relation with 90\% significance.  
}
\label{sosio}
\end{figure}

In the same way, we have fitted the S and Ar to Si ratio  (Figure \ref{sosio}) and derived 
the   S/Si and Ar/Si ratio of  SN Ia ((S/Si)$_{\rm SN Ia}$ and
(Ar/Si)$_{\rm SN Ia}$) to be $\sim $ 1.5 solar and 1.3 solar,
respectively (Table \ref{sn1}).
In contrast to the Fe/Si ratio, W7 and WDDs give nearly the same values
for the ratio between the intermediate elements, S/Si and Ar/Si, and 
the observed ratios are slightly larger than the theoretical values.

\begin{table}
\caption{The abundance pattern of SN Ia obtained from the ICM of M~87
 and those from the nucleo synthesis model of W7 (Nomoto et al. 1984) and
 WDDs (Iwamoto et al. 1999). }
\label{sn1}
\begin{center}
  \begin{tabular}{cccccc}
          & M~87  SNI     & W7 & WDD1 & WDD2& WDD3\\
  S/Si    &1.57$\pm0.32$ &    1.05& 1.14& 1.14 & 1.14\\
  Ar/Si   &1.25$\pm0.45$ &    0.76& 0.89& 0.91 & 0.93 \\
  Fe/Si   &1.10$\pm0.25$ &    2.61& 1.34 & 2.07& 2.99\\
 \end{tabular}
\end{center}
\end{table}

\subsection{Diversity of abundance pattern in SN Ia?}
The light curves of observed SN Ia are not identical but display a
considerable variation (e.g. Hamuy et al. 1996).
In SN Ia, the mass of synthesized Ni$^{56}$ determines the
luminosity of each SN.
Since the mass of the progenitor  should be constant  at 1.4 $M_{\odot}$,  the ratio of mass
of intermediate group elements from Si to Ca, to the mass of Fe and Ni, should depend
on the luminosity of SN Ia.
The observed  luminosity of SN Ia correlates with the type of the host
galaxy, and is
suggested to be related to the age of the system; SNe Ia in  old stellar
system may have smaller luminosities (Iwanov et al. 2000), and hence are suggested to yield a
smaller Fe/Si ratio (e.g. Umeda et al. 1999).

As discussed  in Finoguenov et al. (2002),  the smaller Fe/Si ratio observed
for the ICM around M~87 may reflect the fact that M~87 is an old stellar system.
When we compare the SN Ia abundance pattern for the central and outer
regions of M~87, the outer zone SN Ia abundance pattern depends more
heavily on the adopted abundance pattern of SN II.
As in Finoguenov et al. (2002), we use the SN II pattern derived in
Iwamoto et al. (1999)  for which a Salpeter IMF was adopted.
As indicated in Figure 17, the (Fe/Si)$_{\rm SN Ia}$ of the outer
region is determined to be $\sim$ 1.7 solar, while the central region
data yield a value of $\sim$ 1.2 solar  (Figure \ref{stars}). 
In contrast, when we adopt the abundance pattern of metal poor Galactic
stars as a SN II pattern,   the difference of (Fe/Si)$_{\rm SN Ia}$ of the central and outer regions
becomes smaller.
The average products of SN II may also differ between the
ICM and late-type galaxies, since the IMF of stars may
differ  between early-type and late-type galaxies.
Therefore,   (Fe/Si)$_{\rm SNIa}$ of the outer region still has  some
uncertainty, while the central value is determined to
be $\sim1.2$ solar, which is not affected by the adopted SN II pattern very much.

However, 
the metal poor stars, i.e. those with lower Fe/O, tend to be located
around  larger (Fe/Si)$_{\rm SNIa}$  values (Figure \ref{stars}).
This result suggests that 
the SN Ia were dominated by those with  larger (Fe/Si)$_{\rm SNIa}$ when the
Galaxy was a young stellar system.
Furthermore, based on the ASCA survey of the Virgo cluster, 
Shibata et al. (2002) discovered that the Fe/Si ratio becomes larger at
$R>30'$, where the SN Ia products are accumulated from high $z$.
Although the O abundance cannot be measured, this result may reflect the
fact that the Fe/Si ratio in SN Ia depends on the age of the system.

 The ICM in the  central and outer regions should have a different origin of
metals.  As discussed in Paper I, considering the gas mass and stellar
 mass loss rate,  
most of  the Si and Fe at the center come from present SN Ia in M~87 in the last few Gyr.
In contrast, in the outer regions, metals from SN Ia in M~87 are accumulated
over a longer time scale, and  in addition, other galaxies should
contribute to the metals.
Since the SN Ia rate is expected to be much larger in the past
(Renzini et al. 1993), the SN Ia ejecta which occurred in much younger systems should be 
important in the outer region.

The indication that the abundance pattern of SN Ia in the outer region
of M~87 features a larger (Fe/Si)$_{\rm SNIa}$  ratio compared to the
inner region as found above, can nicely interpreted in line with of these
previous findings. The larger (Fe/Si)$_{\rm SNIa}$ ratio originates from
a younger stellar population, so  SN Ia products that have been
produced earlier in the history of M~87 and the Virgo cluster are
more widely distributed in the ICM.

\subsection{Nucleo synthesis from SN II}

We can also constrain the abundance pattern of SN II in the ICM.

Since Mg and O are not synthesized by SN Ia, the Mg/O ratio reflects those
of the products of  SN II.
The observed Mg/O ratio of $\sim$1.25 solar at $r>0.5'$ derived from the line ratio agrees well with that of the
Galactic stars, although
a higher Mg/O ratio is also allowed  within $r<0.5'$ due to
the uncertainties in the temperature component  below 1 keV.
This value is also quite similar to the Mg/O ratio of 1.3$\pm0.2$ solar
of the ISM of an X-ray luminous elliptical galaxy, derived from the RGS data
(Xu et al. 2002).
The central Mg/O reflects the stellar abundance ratio of M~87, while that of outer
region is the ratio of of metals in the ICM which is an accumulation of
old SN II products ejected from galaxies.
Therefore, at least for the Mg/O ratio, there is no observable difference in
nucleosynthesis products
of SN II between the elliptical galaxies, M~87 and NGC 4636, the ICM and the Galaxy.

The observed S/Si ratio decreases when the O/Si ratio increases (Figure 8, \ref{sosio}).
Considering that the observed Fe/Si ratio is  constant, this indicates that S/Si
ratio is anti-correlated to  the O/Fe ratio, that is  the contribution from  SN
II/SN Ia.
From equation (1), using S instead of Fe,
the S/O ratio in ejecta of SN II ((S/O)$_{\rm SNII}$) is obtained from
the  observed relation in Figure \ref{sosio}.
The Si/O ratio of metal poor stars by Clementini et al. (1999) is 
$\sim 0.7$ solar, while the nucleo synthesis model by Iwamoto et al. (1999)
gives the value of 0.9 solar.
Therefore assuming that (Si/O)$_{\rm SNII}$ is less than 1 solar, the
extension of the observed values in Figure 19 gives (S/O)$_{\rm SNII}$  less than 0.5 solar.
In the same way,  (Ar/O)$_{\rm SNII}$ is derived to be less than 1 solar.

 The observed S abundances of Galactic stars are generally  consistent  with the Si
abundances (e.g. Chen et al. 2002, Takeda-Hidai et al. 2002), although S lines are very week. In contrast, 
the observation of type II planetary nebulae yielded a S/O and Ar/O
ratio  to be 0.4$\pm 0.2$ solar and 1.1$\pm 0.5$ solar (Kwitter \& Henry
2001). 
The nucleosynthesis model of SN II by Nomoto et al. (1997) assuming
Salpeter's IMF gives the
S/O ratio of 0.6 solar, while another model by Woosley \& Weaver (1995)
gives a larger ratio of 1 solar.
Thus, there may be some uncertainty in the S/O ratio synthesized by SN
II even in the Galaxy, and it is also not clear that the Galaxy and the
ICM have the same ratio.
The fact that the  S/Si ratio of the ICM observed with ASCA generally decreases toward
outer radius (Finoguenov et al. 2000) supports the low S/Si ratio in SN
II, although ASCA cannot constrain O abundance.
Therefore, it is important to study outer regions of clusters with the XMM-Newton
where SN II may become dominant (e.g. Fukazawa et al. 2000, Finoguenov
et al. 2000).

In summary, the SN II abundance pattern observed from the ICM mostly agrees
well with that of our Galaxy.
Since the abundance pattern of SN II depends on the IMF of
stars,  this consistency is able to provide strong constraints on it.

\subsection{Comparison with the stellar metallicity profile}

\begin{figure}[]
\resizebox{\hsize}{!}{\includegraphics{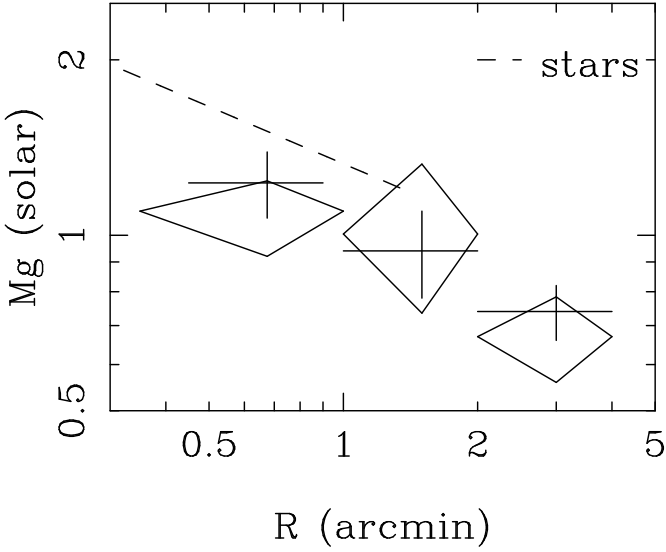}}
\caption{
Mg abundance profile of the EMOS data derived from the spectral fitting
 with the APEC code (diamonds) and from line ratios (crosses). The dashed line represents the stellar metallicity
 derived from Mg$_2$ index (Kobayashi \& Arimoto 2000)
}
\label{ovsmg}
\end{figure}

Figure \ref{ovsmg} shows the observed Mg profile, with the stellar metallicity
profile from the Mg$_2$ index (Kobayashi \& Arimoto 1999).
In addition to the Mg abundance profile derived from the APEC model fit,
that  from the Mg/Si line ratio and deprojected Si abundance
profile is also plotted.
The Mg abundance profile of the ICM has a slightly smaller normalization by
20$\sim$30\% than the stellar metallicity profile at the same radius.

Although the Mg abundance may have some uncertainties due to the Fe-L
atomic data,  the APEC code can well fit the Fe-L/Mg-K structure.
The uncertainty of the Mg abundance due to the uncertainty of the
temperature structure should be also small, at least at $R$=0.5--2$'$, 
since the three temperature model does not change the result, although the
abundances from the two- and single-temperature results differ by a
factor of 1.5--2.
Considering these uncertainties and the difference in observational techniques,
we can conclude that  the Mg abundance of the ICM is consistent with the
stellar metallicity profile at the same radius.
Since we are comparing  abundances in two distinct media, stars and
ISM, which could have very different histories, the abundance results do
not have to agree in general. But this agreement is consistent with the
picture where the central gas in M~87 comes from stellar mass loss as
discussed below.

\subsection{The SN Ia rate of M~87}

The metals in the ICM are a sum of metals ejected from the galaxy and those
contained previously within the ICM.
The observed Fe abundance becomes nearly constant within 2$'$, which indicates 
that the gas is dominated by gas ejected from the galaxy.
The observed Mg and Fe abundances within 0.35--2$'$ are $1.1\pm0.2$ solar and
1.54$\pm0.1$ solar, respectively.
Both the nucleosynthesis model by Nomoto et al. (1997) and observation of
Galactic stars (e.g. Clementini et al. 1999; Nissen et al. 1994) 
indicate that (Fe/Mg)$_{\rm SN II}$ is 0.3--0.5 solar.
 Therefore, within $R<2'$, the Fe abundance synthesized by SN II 
should be 0.3--0.6 solar.
Subtracting it, the Fe abundance synthesized by SN Ia 
is derived to be $\sim$ 0.9--1.4 solar.

The  metallicity of gas ejected from the galaxy are a sum of stellar metallicity
and the contribution from SN Ia, and can be expressed as,
\begin{eqnarray}
\nonumber z^i =\frac{\alpha_*y^i_{*}+\alpha_{\rm SN}y_{\rm
 SN}^i}{\alpha_*+\alpha_{\rm SN}} 
 \sim y_{*}^i+\langle\frac{\alpha_{\rm SN}}{\alpha_{*}} \rangle y_{\rm SN}^i\\ 
 = y_{*}^i+\langle\frac{\theta_{\rm SN} M_{\rm SN}^i}{\alpha_{*}} \rangle
\end{eqnarray}
(Loewenstein and Mathews 1991; Ciotti et al.\ 1991), where $z^i$ is
the mass fraction of the $i$th element, $\alpha_*$ and $\alpha_{\rm
SN}$ are the mass loss rates of stars and SNe respectively, and
$y_{*}^i$ and $y_{\rm SN}^i$ are their yields.
 $\theta_{\rm SN}$ is the SN Ia rate and $M_{\rm SN}^i$ is the mass of the
$i$th element synthesized by one SN Ia.
 Converting to abundance in unit of the solar ratio, Fe abundance  synthesized by SN Ia of
gas ejected from an elliptical galaxy becomes

\begin{eqnarray}
\nonumber {{{A_*^{\rm{Fe}}}_{\rm SN Ia}}}+\left(\frac{\theta_{\rm SN} \rm M_{\rm SN}^{\rm Fe}}{\alpha_*}\right)\left(z^{\rm Fe}_{\rm solar} \right)^{-1} \\
={{{A_*^{\rm{Fe}}}_{\rm SN Ia}}}+2.1 \frac{
{\left(\frac{\theta_{\rm SN}}{1.0\times10^{-13}/{\rm yr}/L_{\rm B}}\right)}
\left(\frac{M^{\rm Fe}_{\rm SN}}{0.4M_\odot}\right)}  
{{\left(\frac{\alpha_*}{1.5\times10^{-11}M_\odot  /{\rm yr}/L_{\rm B}}\right)}}
\end{eqnarray}

 Here, ${{{A_*^{\rm{Fe}}}_{\rm SN Ia}}}$ is stellar Fe abundance
synthesized by SN Ia and $M^{\rm Fe}_{\rm SN}$ is the mass of Fe synthesized by one
SN Ia.
 $z^{\rm Fe}_{\rm solar}$ is the Fe metallicity  of  gas with the solar abundance.
Considering the fact that recent observations  of M 87 and A 496
(Tamura et al. 2001) indicate a smaller Fe/Si
ratio synthesized by SN Ia than usually assumed previously,
 $M_{\rm Fe}$ of one SN Ia should be $\sim 0.4M_\odot$.
From a theoretical stellar evolutionally model for a stellar population, 
the stellar mass loss rate is approximated by $1.5\times10^{-11}L_B
t_{15}^{-1.3} M_\odot /yr$, where $t_{15}$ is the age in unit of 15 Gyr,
and $L_B$ is the B-band luminosity (Ciotti et al. 1991).
An infrared observation, which can directly trace mass loss rate from
asymptotic giant branch stars, gives a consistent value within a factor
of 2 with the theoretical value (Athey et al. 2002).
The SN Ia rate of elliptical galaxies is estimated to be 0.13$\pm$0.05 ${h_{75}}^2$ SN Ia/100yr/10$^{10}L_\odot$
(Cappellaro et al. 1997).
 Here, $h_{75}$ is a Hubble constant in unit of 75 km s$^{-1}$ Mpc$^{-1}$.
Using the Hubble constant of 72$\pm8$ km s$^{-1}$ Mpc$^{-1}$ (Freedman
et al. 2001), the present contribution from SN Ia to the Fe abundance becomes
1.4--3.6 solar.

Considering the large error  due to a statistical error
and uncertainties in bias corrections of SN Ia from optical observation,
this value is consistent with 
the  Fe abundance synthesized by SN Ia within 2$'$, 0.9--1.4
solar, which is a sum of ejecta of present SN Ia and that
was trapped in stars and comes from stellar mass loss.
The contribution from stars to Fe synthesized by SN Ia may be small, 
since stars in giant elliptical galaxies are
thought to have a SN II like abundance pattern (e.g. Worthey et al. 1993).

\subsection{Implication of S abundance}

 In Sec.7.5 we  indicated that the S/Si ratio in SN II is much lower than 1 solar.
If this fact is  valid universally in  the ICM, then
the S/Si ratio  may be a better indicator for the contribution from the SN Ia/SN II than
the Fe/Si ratio.
Of course, the best indicators of  SN II are  O, Ne and Mg, which are
not synthesized in SN Ia.
However, emission lines of Ne and Mg are often hidden in the strong
Fe-L lines, and the O abundance can be measured only in relatively cool clusters.

As discussed in Sec.7.4 the Fe/Si ratio of SN Ia may be expected to show a variation, since the luminosity
of the SN Ia should reflect the amount of Ni$^{56}$ synthesized by SN Ia.
The pure SN II ejecta may provide the smaller Fe/Si ratio.
Then SN Ia in young systems occur, which may have a larger Fe/Si ratio and
so the Fe/Si ratio increases.
Finally, SN Ia in  old systems decrease the Fe/Si ratio again.
In contrast, at least in  Iwamoto et al. (1999), the S/Si ratio in the
various SN Ia is almost constant.
In addition, as shown in Sec. 5.2, the S/Si line ratio does not depend on the
plasma temperature, while  that of the Fe/Si is a very strong
function. This means that, the derived S/Si does not depend on
the temperature structure of the ICM.  This is a very important
advantage for the cool cores of
the cluster center, where many temperature components exists.

The problem is that both Si and S are synthesized in the same nucleo
process, i. e. in the same region in the SN.
Therefore, it is surprising that the S/Si ratio shows a radial variation.
Unfortunately, the S abundance measured  of the Galactic stars has very
large uncertainties, and
the S/Si ratio of the SN II may also depend on the IMF of stars.
Therefore, it is quite important to calibrate the S/Si vs. O/Fe pattern 
in luminous clusters.

\section{Summary and Conclusion}

Based on
 the temperature structure studied in Paper I,  abundance
profiles  of  7 elements of the ICM around M~87 are obtained using
the deprojected spectra.
We have discussed the use of one- and two-temperature models in the fit
 and the application of the MEKAL and APEC codes. Two temperature models
 are only important in the region, R$<2$ arcmin. The previous problems
 with an earlier version of the APEC code are now resolved in the
 present version and this code provides now better fitting solutions in
 general. In addition, using line ratio profiles of the projected and deprojected data,
the abundance ratio profiles are obtained with smaller uncertainties,
 considering the temperature dependence of the line ratio,  
 The main results obtained are as follows,

\begin{itemize}
 \item Abundance profiles of 7 elements are obtained with high accuracy
       (e.g. $\delta$Fe$\sim\delta$Si$\sim$5\%,  $\delta$O$\sim$10\%).
\item       The Si and Fe abundance profiles have steep gradients at $>2'$ 
       with a Fe/Si ratio of $0.9\pm0.1$  solar, and become constant
       within this radius at 1.5--1.7 solar.
\item       The S/Si ratio is about 1 solar at $r<$ 4$'$ and decreases to
       $\sim$ 0.7 solar at $r>10'$.
\item       The Ar/Si and Ca/Si are about 0.8 solar and 1.5 solar,
       respectively, although errors are relatively large.
\item  The O and Mg abundances are smaller than the abundances of Fe and the  intermediate
       elements.  The O/Si ratio is less than  half solar at the center and
       increases with radius.
       The Mg/O ratio is 1.25 solar and consistent with no radial
       gradients at least at $r>0.5'$.
 \item The observed abundance pattern among O, Mg, Si, Ca and Fe of the ICM is located at an extension of
that of Galactic stars. Therefore, the abundance pattern of the ICM is
       not peculiar and it should strongly constrain the products of SN
       Ia and SN II of early-type and late-type galaxies.
\item  The Si/O/Fe diagnostics shows large Si contributions by SN Ia,
       which may reflect the observational finding that SN Ia in old stellar systems are fainter.
\item  The observed Mg abundance of the ICM is consistent with
        the stellar metallicity profile from the Mg$_2$ index at the same radius.
This result is consistent with stellar mass loss as the source of the
       gas in the very central region of M~87.
\item  The central abundance can be explained with the observed SN Ia rate and
       SN Ia model yields.
\item  The different radial profiles between Si and S suggest that the
       S/Si ratio synthesized from SN II is much smaller than the solar
       ratio. Then, the S/Si ratio may be a  better indicator of the relative
       contribution from SN Ia and SN II, when O and Mg abundances
       cannot  be observed.
\end{itemize}

We are actually very fortunate to have M~87 and the Virgo cluster as the
closest galaxy cluster in our neighborhood, since it offers an ICM in
just this low temperature range where the spectra are richest in
spectral lines for this type of abundance diagnostics.
Therefore it would be worthwhile to use this unique case for even deeper
observations and more detailed investigations by X-ray spectroscopy in
the future.

\begin{acknowledgements}

We would like to thank Nobuo Arimoto and Kuniaki Masai for valuable
 suggestions on this work. 
 This work was supported by the Japan
 Society for the Promotion of Science (JSPS) through its Postdoctoral
 Fellowship for Research Abroad and Research Fellowships for Young
 Scientists.

 The paper is based on observations obtained with XMM-Newton, an ESA science mission with instruments and contributions direct
by funded by ESA Member States and the USA (NASA).
 The XMM-Newton project is supported by the Bundesministerium f\"ur Bildung und Forschung, Deutsches Zentrum  f\"ur Luft und R
aumfahrt (BMBF/DLR), the Max-Planck Society and the Haidenhain-Stiftung.

\end{acknowledgements}

 \end{document}